\documentclass[aps, prx, reprint, superscriptaddress]{revtex4-2}
\usepackage[dvips]{graphicx}
\usepackage{amsmath,amssymb,amsthm,mathrsfs,amsfonts,dsfont,amscd,keyval}
\usepackage{mathtools,mathrsfs,esvect}
\usepackage{yfonts} %  for \vcentcolon
\usepackage{ytableau}
\usepackage{graphicx}
\usepackage{braket}
\usepackage{bm}
\usepackage{fancyhdr}
\usepackage{enumerate}
\usepackage{color}
\usepackage{hyperref}
\usepackage{tabularx}
\usepackage{times}
\hypersetup{colorlinks=true, linkcolor=purple, citecolor=purple, urlcolor=black}
\usepackage{multirow}
\usepackage{float}
\usepackage{tensor}
\usepackage{multirow}
\usepackage{physics}

\usepackage{xcolor}
\usepackage{listings}
\usepackage{tcolorbox}
\usepackage{fancyvrb}
\usepackage{booktabs}
\usepackage{enumitem} 
\usepackage{IEEEtrantools}

\usepackage{tikz} % TikZ package for graphics
\usetikzlibrary{quantikz} % Quantikz library for quantum circuit diagrams

\usepackage[utf8]{inputenc}
\usepackage[T1]{fontenc}

\usepackage{hyperref}% add hypertext capabilities
\hypersetup{
  colorlinks   = true, %Colours links instead of ugly boxes
  urlcolor     = blue, %Colour for external hyperlinks
  linkcolor    = blue, %Colour of internal links
  citecolor   = blue %Colour of citations
}

\lstdefinestyle{quantumcode}{
    basicstyle=\ttfamily\small\color{white},
    backgroundcolor=\color{black!90},
    numbers=left,
    numberstyle=\tiny\color{gray},
    numbersep=8pt,
    frame=tb,
    framerule=0.5pt,
    rulecolor=\color{gray},
    tabsize=4,
    breaklines=true,
    postbreak=\mbox{\textcolor{green}{$\hookrightarrow$}\space},
    keywordstyle=\color{green!70!black},
    commentstyle=\color{cyan},
    stringstyle=\color{white},
    morekeywords={QuantumCircuit, append, h, measure, range}, % 添加量子计算关键词
    literate=* 
        {:}{{\textcolor{white}{:}}}{1} % 冒号保持白色
        {=}{{\textcolor{white}{=}}}{1}
        {(}{{\textcolor{white}{(}}}{1}
        {)}{{\textcolor{white}{)}}}{1}
        {*}{{\textcolor{white}{*}}}{1}
}

\renewcommand\vec{\mathbf}

\newcommand{\Var}{\mathrm{Var}}

\theoremstyle{definition}
\newtheorem{definition}{Definition}[section]
\newtheorem*{example}{Example}

\theoremstyle{plain}
\newtheorem{theorem}[definition]{Theorem}

\theoremstyle{remark}

\newcommand{\comments}[1]{}
\makeatletter
\def\l@subsubsection#1#2{} % disable subsubsections in the TOC
\makeatother

\begin{document}

\let\oldaddcontentsline\addcontentsline

\renewcommand{\addcontentsline}[3]{}

\title{Provably Efficient Quantum Algorithms for Solving Nonlinear Differential Equations Using Multiple Bosonic Modes Coupled with Qubits}

\author{Yu Gan}
\altaffiliation[These authors ]{contributed equally to this work.}
\affiliation{%
Computer Science, University of Pittsburgh, Pittsburgh, PA 15261, USA
}%

\author{Hirad Alipanah}
\altaffiliation[These authors ]{contributed equally to this work.}
\affiliation{%
  Mechanical Engineering and Materials Science, University of Pittsburgh, Pittsburgh, PA 15261, USA
}%

\author{Jinglei Cheng}
\affiliation{%
 Computer Science, University of Pittsburgh, Pittsburgh, PA 15261, USA
}%

\author{Zeguan Wu}
\affiliation{%
Computer Science, University of Pittsburgh, Pittsburgh, PA 15261, USA
}%

\author{Guangyi Li}
\affiliation{%
Computer Science, University of Pittsburgh, Pittsburgh, PA 15261, USA
}%
\affiliation{%
Electrical Engineering and Computer Science, University of Michigan, Ann Arbor, MI
48109, USA
}%

\author{Juan~Jos\'e~Mendoza\nobreakdash-Arenas}
\affiliation{%
 Mechanical Engineering and Materials Science, University of Pittsburgh, Pittsburgh, PA 15261, USA
}%
\affiliation{%
 Physics and Astronomy, University of Pittsburgh, Pittsburgh, PA 15261, USA
}%

\author{Peyman Givi}
\affiliation{%
Mechanical Engineering and Materials Science, University of Pittsburgh, Pittsburgh, PA 15261, USA
}%
\affiliation{%
Petroleum Engineering, University of Pittsburgh, Pittsburgh, PA 15261, USA
}%

\author{Mujeeb R. Malik}
\affiliation{%
 NASA Langley Research Center, Hampton, VA 23681, USA
}%

\author{Brian J. McDermott}
\affiliation{%
 Naval Nuclear Laboratory, Schenectady, NY 12301, USA
}%

\author{Junyu Liu}
\email{junyuliu@pitt.edu}
\affiliation{%
Computer Science, University of Pittsburgh, Pittsburgh, PA 15261, USA
}%

\date{\today}
\maketitle

\noindent{\bf Quantum computers have long been expected to efficiently solve complex classical differential equations. Most digital, fault-tolerant approaches use Carleman linearization to map nonlinear systems to linear ones and then apply quantum linear-system solvers. However, provable speedups typically require digital truncation and full fault tolerance, rendering such linearization approaches challenging to implement on current hardware. Here we present an analog, continuous-variable algorithm based on coupled bosonic modes with qubit-based adaptive measurements that avoids Hilbert-space digitization. This method encodes classical fields as coherent states and, via Kraus-channel composition derived from the Koopman-von Neumann (KvN) formalism, maps nonlinear evolution to linear dynamics. Unlike many analog schemes, the  algorithm is provably efficient: advancing a first-order, $L$-grid point, $d$-dimensional, order-$K$ spatial-derivative, degree-$r$ polynomial-nonlinearity, strongly dissipative partial differential equations (PDEs) for $T$ time steps costs $\mathcal{O}\left(T(\log L + d r \log K)\right)$. The capability of the scheme is demonstrated by using it to simulate the one-dimensional Burgers' equation and two-dimensional Fisher-KPP equation. The resilience of the method to photon loss is shown under strong-dissipation conditions and an analytic counterterm is derived that systematically cancels the dominant, experimentally calibrated noise. This work establishes a continuous-variable framework for simulating nonlinear systems and identifies  a viable pathway toward practical quantum speedup on near-term analog hardware.
}

\begin{figure*}[t]
    \centering
    \includegraphics[width=\linewidth]{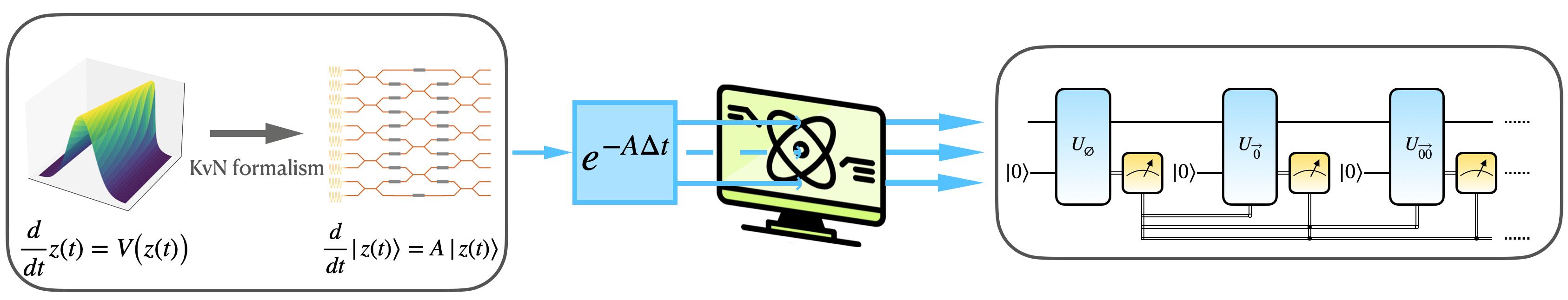}
    \caption{Overview of the continuous-variable quantum algorithm.
    \emph{Left:} A nonlinear field evolution $\dot z=V\left(z(t)\right)$ is lifted by the KvN map to a linear operation generated by $A$ (which encodes the differential equation information) on an enlarged space.
    \emph{Middle:} Each short step is implemented as a local CPTP map $K_a=e^{-A\Delta t}$ acting on bosonic modes prepared in multimode coherent states that encode the classical field.
    \emph{Right:} $K_a$ is compiled into a measurement-adaptive binary-tree circuit: a system–ancilla unitary $U$, measurement and reset at each layer, and post-selection on $\ket{0}$ yield logarithmic circuit depth in the Kraus rank.}
    \label{fig:Overall_diagram}
\end{figure*}

\section{Introduction}
Quantum computing, as suggested by Richard Feynman \cite{feynman2018simulating}, is naturally suitable for quantum physics–related problems such as  chemical systems and many-body dynamics \cite{lloyd1996universal}. Quantum algorithms have demonstrated theoretical speedups over their classical counterparts for problems such as integer factorization \cite{Shor_1997}, unstructured search \cite{grover1996fast}, and linear system solving \cite{harrow2009quantum}. Some of the use cases of quantum algorithms are designed not only for quantum systems but also for classical  problems \cite{givi2020quantum}.  Although significant progress has been made on algorithm development, many of those algorithms require fully fault-tolerant digital architectures that are not yet available.\\

Numerical simulation of differential equations is of significant interest in many fields, not only in quantum  but also in classical systems such as fluid dynamics, to economics, finance, and more  \cite{leyton2008quantum,berry2014high,Liu_2021,kiani2022quantum,childs2020quantum,xin2020quantum,childs2021high,kyriienko2021solving,an2022theory,Jin_2024,liu2023efficient,meng2023quantum,chen2024enabling,jin2024quantum,Jin_2024112707,liu2024towards,lu2024quantum,meng2024quantum,meng2024simulating,su2024quantum,ye2024hybrid,alipanah2025quantum,costa2025further,fang2025qubit,golse2025quantum,meng2025geometric,shang2025designing,su2025efficient,tennie2025quantum,wang2025quantum,wang2025simulating,wang2025towards,yang2025quantum,zhang2025data,zhu2025quantum}. In computational fluid dynamics (CFD), the complexity of the flow increases as the  Reynolds number $Re$ increases.     In classical direct numerical simulation (DNS), the number of grid points needed to resolve all length and time scales increases as $\mathcal{O}(Re^{9/4})$ \cite{LNBG20,Wang2015TowardsHA}. As in quantum many-body physics, simulating turbulence suffers from rapid growth in the effective system size. \\

Most of the quantum algorithms proposed with rigorous speedup are based on quantum linear system solvers, such as the Harrow–Hassidim–Lloyd (HHL) algorithm \cite{harrow2009quantum} and its variants. These  solvers typically employ Carleman linearization~\citep{Carleman_1932} to embed nonlinear dynamics into a truncated high-dimensional linear system, followed by a quantum linear solver to simulate its evolution~\citep{Liu_2021}. To avoid the ``dimensional explosion" in the truncated linear system, there must be a limit on the nonlinearity of the dynamics \cite{Liu_2021}. These HHL-based  quantum algorithms demand fully fault-tolerant digital systems requiring  extra overhead in qubit count, circuit depth, and state preparation. Such requirements make linearization-based quantum algorithms impractical for current quantum devices, and the additional challenge of controlling truncation errors further limits their feasibility.\\

Contemporaneous with Carleman’s proposal of linearization for nonlinear systems, Koopman and von Neumann \cite{Koopman1931,vonNeumann1932} developed an abstract functional-analytic interpretation for nonlinear equations. The method is known as the Koopman–von Neumann (KvN) formalism by which a nonlinear equation can be associated with infinitely many linear equations embedded in an artificial Hilbert space. 
This emergent Hilbert space neither complies with Born's rule nor exhibits entanglement as in quantum physics. Instead, it provides an efficient, quantum-like mathematical description of the original nonlinear dynamics, whether quantum or classical. Thus, Carleman’s linearization approaches can be understood as a finite-dimensional digital truncation of this Hilbert space.  (This truncation puts a limit on the extent of nonlinearity that can be dealt with by Carleman’s linearization). A somewhat similar approach is followed in machine learning through the kernel method and support vector machines, which linearize nonlinear data into a higher-dimensional feature space \cite{mohri2018foundations}.\\

In this work, it is shown that the KvN Hilbert space can be generated by multiple bosonic modes, i.e. multiple products of Fock spaces, where the time evolution in differential equations is represented by the dynamics of coherent states. This mapping enables the development of new quantum algorithms in continuous-variable quantum devices.  The workflow of the algorithm is depicted  in Fig.~\ref{fig:Overall_diagram}, and the procedure goes as follows: the initial condition profiles are uploaded  onto coherent states.  These states evolve through quantum control following the target differential equations, concluded with the measurement of the  annihilation operators to extract the solution.   The  algorithm also provides  a scaling advantage. For highly dissipative equations, time evolution can be treated as dissipative open-system dynamics in multiple bosonic modes.  This dynamics is described as a quantum channel or a completely positive trace-preserving (CPTP) map with post-selection for some Kraus rank. Quantum control theory established in circuit quantum electrodynamics (circuit QED)~\cite{lloyd2001engineering,Shen_2017,ma2021quantum} shows that for Kraus rank-$N$, only  $O(\log N)$ circuit depth is needed to construct the corresponding quantum channel via couplings between bosonic modes and qubits. For ordinary differential equations (ODEs), the Kraus rank is upper-bounded by the number of functions to be solved, and for partial differential equations (PDEs), it is upper-bounded by the number of grid points in spatial discretization.
Since large-scale simulations of differential equations such as the Navier-Stokes~\cite{givi2020quantum} require of  order of billions of grid points, algorithms with logarithmic scaling in circuit depth provide practical benefits.  The  scaling observed here  is also consistent with the results in Ref.~\cite{Liu_2021}. The tree-like structure of quantum control (Fig.~\ref{fig:binarytree_combined}) can be understood as a continuous-variable counterpart of conditional rotations in HHL or the tree-like architecture in quantum random access memories (QRAMs)~\cite{giovannetti2008quantum}. \\

The bosonic space naturally avoids digital truncation and truncation errors, and significantly reduces the overheads of HHL. Beyond circuit QED, the proposed algorithm may also be implemented in photonics \cite{tiecke2014nanophotonic}, ion traps \cite{leibfried2003quantum}, or neutral atom arrays \cite{signoles2014confined}, where bosonic modes with photons can host hundreds of energy levels with strong nonlinearity in emergent Hilbert space. Although fault tolerance remains challenging in analog, bosonic, continuous-variable setups \cite{menicucci2014fault}, we argue that for strongly dissipative systems and some noise models there exists natural noise resilience. Some quantum noise channels, appearing as additional terms in differential equations (e.g., photon loss), can be suppressed during long-time evolution, making the solutions robust. In addition, an error mitigation technique is developed to reduce the effect of errors by introducing counterterms. Unlike  existing proposals, the  method developed here can deal with arbitrarily large nonlinearities at a linearly increasing cost. 
\textit{The algorithm  also provides some insights into why some of the quantum computing algorithms can efficiently solve highly nonlinear classical problems. These problems suggest  the emergence of an artificial Hilbert space from the KvN formalism. Some quantum computing algorithms can  mimic this emergent space using the real, physical Hilbert space of quantum mechanics \cite{havlivcek2019supervised,succi2025foundational}.}

\section{Koopman-von Neumann formalism for quantum information}
\label{sec:kvn}
The Koopman--von Neumann (KvN) formalism trades nonlinearity in the original variables for linear evolution in an enlarged Hilbert space. Recall the properties of a bosonic mode with annihilation operator $a$ ( $[a,a^\dagger]=1$). One can define the unnormalized coherent state as $\ket{z} = e^{z a^\dagger}\ket{0}$. The state satisfies $a\ket{z} = z\ket{z}$ and $a^\dagger \ket{z} = \frac{\partial}{\partial z}\ket{z}$, which are known as Segal–Bargmann identities.  The derivation is provided in the Supplementary Materials~\ref{subsection:deri-coherets}. 
For the scalar ODE \(\frac{d}{dt}z(t)=V \big(z(t)\big)\),
one  can encode the evolution of $z$ into the coherent state variable. Using  the chain rule of differentiation, it is straightforward to show that the evolution of the resulting coherent state is
$\frac{d}{dt}|z(t)\rangle=  a^\dagger|z(t)\rangle \cdot \, V(z(t)).$
Applying the Segal–Bargmann identities,  the operator $V$ can be re-written such that
$V(a)\,|z\rangle := V(z)\,|z\rangle.$ Defining an effective operator $A := a^\dagger V(a),$
the original nonlinear ODE is mapped into the  linear evolution:

\begin{equation}
\frac{d}{dt}|z(t)\rangle = A\,|z(t)\rangle .
\label{eq: lifted evolution}
\end{equation}
The trajectory $z(t)$ can be recovered using
\begin{equation}
  z(t)=\frac{\langle z(t)|\,a\,|z(t)\rangle}{\langle z(t)|z(t)\rangle},
  \label{eq:expval}
\end{equation}
which follows directly from the Segal–Bargmann identities. In the model developed here,  the nonlinear flow in $z$ is encoded into the amplitudes of the coherent state $\ket{z}$ and the nonlinear evolution of $z$ is mapped to a linear evolution for $\ket{z}$.
 The method can be generalized to the multivariate case. 
Let $z(t)=\left(z_1(t), \ldots, z_n(t)\right)$. The corresponding multimode coherent state is the tensor product $|z(t)\rangle=\bigotimes_i\left|z_i(t)\right\rangle$, and each $z_i(t)$ is governed by a corresponding nonlinear ODE.  The general coherent state is now defined as the tensor product of the coherent states corresponding to each $z_i(t)$. It is easy to verify that the bosonic mode property and the Segal-Bargmann identities can be extended to the multivariate case, which can be constructed with  the transformed state $\ket{z(t)}$.  As in the univariate case,  an effective operator $A(t)$ is defined for the original vector field operator. Expanding the original vector field operator corresponding to $z_i(t)$ in monomials with coefficients \(c_{j,k_1\dots  k_m}(t)\) gives the normal-order generator:
\begin{equation}
\label{eq:kvn-A-series}
A(t)=\sum_{j=1}^{n} a_j^\dagger \left(\;\sum_{m=0}^{\infty}\ \sum_{k_1,\dots ,k_m=1}^{n}
c_{j,k_1\dots  k_m}(t)\; a_{k_1}\cdots a_{k_m}\right).
\end{equation}
For a polynomial vector field with nonlinearity of degree \(r\), the inner sum truncates at \(m\le r\).
With the KvN formalism, the transformed state $\ket{z(t)}$ evolves linearly under the operator $A(t)$.
Moreover, by projecting the transformed state back using Eq.~\eqref{eq:expval},  the original dynamics is recovered. A detailed description is provided in the Supplementary Materials~\ref{subsection:reco-clsskvn}.\\

There exists a  transparent link between our method and Carleman linearization. When the differential operator at the grid point $j$, denoted by $F_j$, has a polynomial nonlinearity of degree $r$, we see that $F_j(a)$ contains $r$ annihilators and $a_j^\dagger$ increases the degree by one. Consequently, $A$ couples number-state sectors of total degree $m$ only to degrees up to $m+(r-1)$. 
This is the same ``degree bandwidth’’ that appears when stacking monomials $(z,z^{\otimes2},z^{\otimes3},\dots )$ in Carleman’s construction to obtain a linear ODE \cite{bellman1997introduction}.
Projecting the KvN evolution so that it includes only polynomial terms up to degree $K$ gives the same result as
a degree-$K$ Carleman truncation, while the degree-1 block read out via Eq.~\eqref{eq:expval} reproduces the solution to the original dynamics. In the next sections, we will introduce how to use the Kraus representation and quantum channel to solve the PDE with the KvN formulation. In the Supplementary Materials~\ref{appsec:lid driven}, we discuss the alternative way to solve PDE in KvN formulation using Trotterization method and show that it leads to the same forward Euler equation.

\section{Kraus Representation}\label{sec:Kraus}

According to Choi’s theorem, any completely positive trace-preserving (CPTP) map from Hilbert space $\mathcal{H}_A$ to Hilbert space $\mathcal{H}_B$, $\mathcal{N}: \mathcal{L}(\mathcal{H}_A) \rightarrow \mathcal{L}(\mathcal{H}_B)$, admits a Kraus decomposition $\mathcal{N}(X) = \sum_{b\in\mathcal{B}} K_b\,X\,K_b^{\dagger},\ \forall X \in \mathcal{L}(\mathcal{H}_A)$
with Kraus operators satisfying $\sum_{b\in\mathcal{B}} K_b^{\dagger} K_b = \mathbb{I}.$ To implement Kraus operators, we couple the bosonic register with a single ancilla qubit initialized in $\ket{0}$. Then, we apply a joint unitary to the coupled system and perform a projective measurement on the ancilla qubit. 
When the outcome is $b\in\mathcal{B}$, the system undergoes the evolution governed by $K_b$. We designate a post-selected ``successful" branch $a\in\mathcal{B}$ (outcome $\ket{0}$), parametrized as
\begin{equation}
    K_a = e^{-A\,\Delta t},\quad\  
    \sum_{b\neq a} K_b^{\dagger} K_b = \mathbb{I} - K_a^{\dagger} K_a ,
\label{eq:krausmodel}
\end{equation}
with $A\succeq 0$ a positive semidefinite generator. In the simplest nontrivial case $|\mathcal{B}|=2$, this reduces to 
$K_a = e^{-A\Delta t}$ and $K_{\bar a} = \sqrt{\mathbb{I}-K_a^\dagger K_a}$. 
For general rank-$N$ channels ($|\mathcal{B}| = N$), the other branches $\{K_b\}_{b\neq a}$ are supplied recursively by QR/SVD completion inside a binary-tree structure.\\

\begin{figure*}[t]
    \centering
    \includegraphics[width=0.48\textwidth]{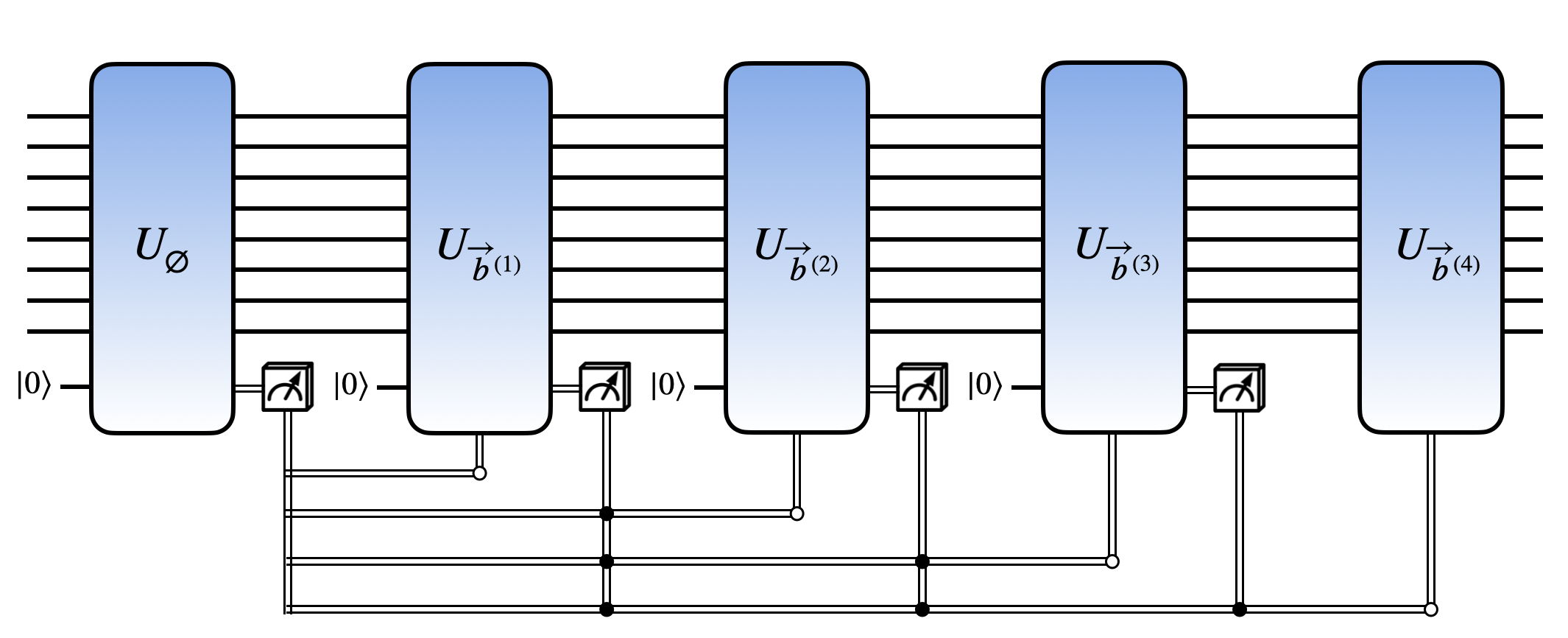}\hfill
    \includegraphics[width=0.48\textwidth]{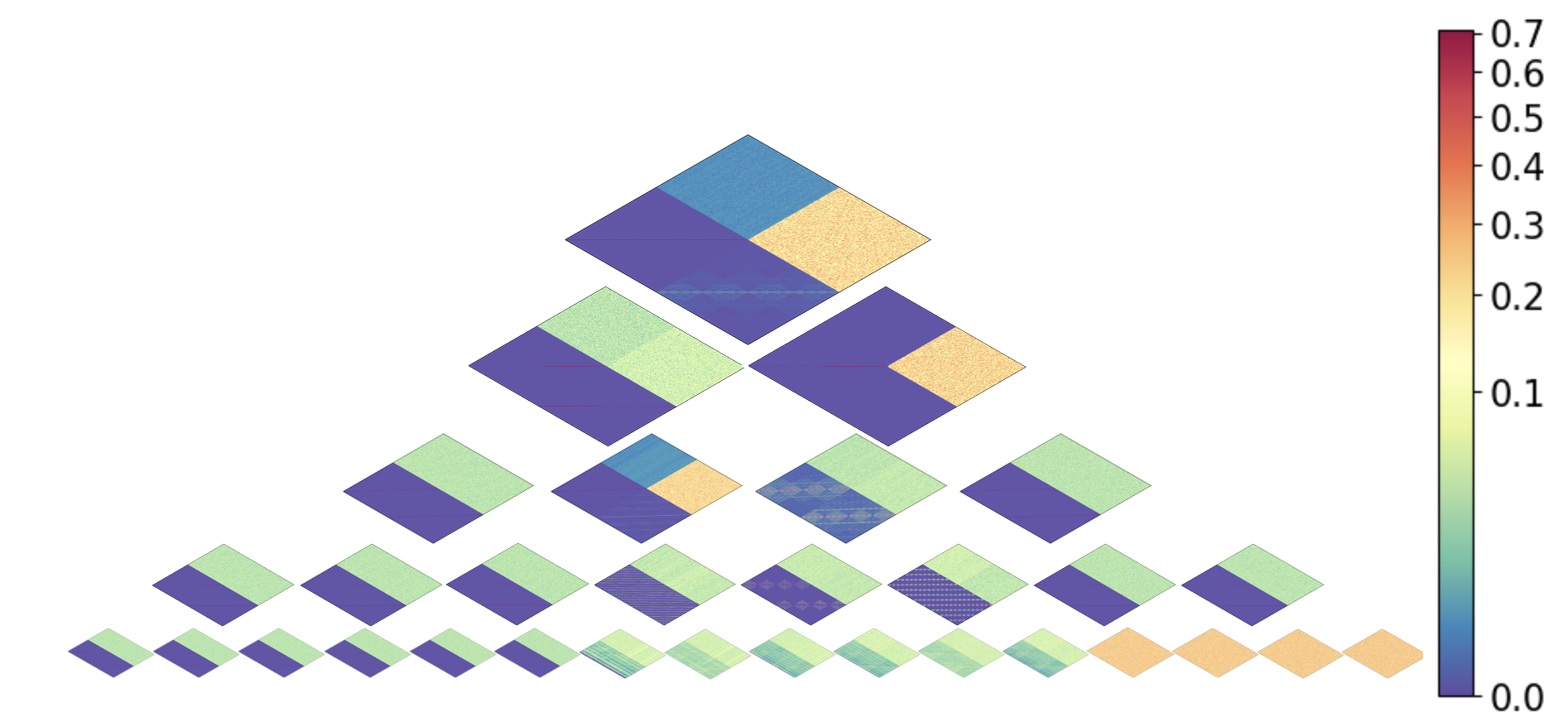}
    \caption{Circuit realization and unitary construction for a 1D Burgers' equation.
    \emph{Left:} Measurement–adaptive binary-tree realization of an $N$-rank-Kraus-operator CPTP map using a single ancilla qubit~\citep{Shen_2017}. At depth $\ell$, the joint unitary $U_{\vec{b}^{(\ell-1)}}$ is applied, where $\vec{b}^{(\ell-1)}$ records the past measurement outcomes. 
    After each interaction, the ancilla is measured and reset. Post-selection on $\ket{0}$ (empty-circle control) singles out the Kraus operator $K_a$.  
    \emph{Right:} Magnitude plots of all compiled block unitaries \(U_{\vec{b}}\) for a four-mode bosonic system truncated to four levels per mode
    ($d_{\mathrm{sys}}=256$), arranged hierarchically from the root ($U_{\varnothing}$) to depth-4 leaves ($U_{0000}, \dots, U_{1111}$ with subscripts in the lexicographical order). 
    For each \(U_{\vec{b}}\), the left half encodes the analytically derived Kraus block \(\langle b_{\ell+1} | U_{\vec{b}} | 0 \rangle\),
    while the right half shows the QR-completed orthonormal complement required to form a valid unitary.}
    \label{fig:binarytree_combined}
\end{figure*}

To compile this Kraus-decomposition-based postselection-involved dynamics into a shallow circuit, we adopt the binary-tree construction from ~\citep{lloyd2001engineering,Shen_2017}. 
A CPTP map of Kraus rank $N$ is decomposed into $\lceil \log_2 N \rceil$ adaptive rounds of ancilla interaction, measurement, and feed-forward. In round $l$, a joint system–ancilla unitary $U_{\vec b^{(\ell-1)}}$ is applied, where $\vec b^{(\ell-1)}=(b_1,\dots,b_{\ell-1})$ records the results of all previous ancilla measurements. The tree therefore branches adaptively: the root corresponds to $U_\varnothing$, the next layer selects $U_0$ or $U_1$ depending on $b_1$, and so on. 
This architecture, illustrated in the left side  of Fig.~\ref{fig:binarytree_combined}, realizes depth $\mathcal{O}(\log N)$, in contrast to a textbook Stinespring dilation which in general requires an environment of dimension $d^2$ (corresponding to $N=d^2$) and a joint unitary in $SU(d^3)$ for a $d$-dimensional system~\citep{Nielsen_Chuang_2010}. In our setting,  we do not compile the full channel but instead focus on a single post-selected trajectory: only branches with $b_\ell=0$ are retained (post-selection on $\ket{0}$), directly enforcing the target Kraus operator $K_a$. This selective evolution is particularly natural in circuit-QED platforms, where ancilla reset and high-fidelity measurement are available in each round. A concrete demonstration is given on the right side  of Fig.~\ref{fig:binarytree_combined}. This explicit construction verifies that the binary-tree procedure yields a complete set of channels satisfying the Kraus consistency condition.

\section{Quantum Channel Execution and Sampling Overhead}\label{sec:channel}

We implement the short-time Kraus channel using measurement-conditioned unitaries and separate the spatial and temporal contributions to the cost. At each step $\Delta t$, the system–ancilla interaction realizes an $N$-outcome map 
$\rho \mapsto \sum_{b\in\mathcal{B}} K_b \rho K_b^\dagger$
with the short-time parameterization of Eq.~\eqref{eq:krausmodel},
where $\rho$ denotes the system’s density operator, representing its quantum state.
Post-selection on a designated outcome $a\in\mathcal{B}$ (ancilla state $\ket{0}$) enforces $K_a$ with success probability
\begin{equation}
p_a \;=\; \operatorname{Tr} \big(K_a^\dagger K_a \,\rho\big)
    \;\approx\; 1 - 2 \Delta t \cdot \operatorname{Tr} \big(A \rho\big) + \mathcal{O}(\Delta t^2).
\end{equation}
To ensure that the outcome of the ancillary measurement corresponds to \( K_a \) with probability $p_a$ greater than \( 1 - \epsilon \), it suffices to choose a time step \( \Delta t \) that satisfies the following:
\begin{equation}
\label{eq:stepsize-constraint}
\Delta t \;\le\; \frac{\epsilon}{2\,\operatorname{Tr} \big(A \rho\big)}.
\end{equation}
The value $\operatorname{Tr}(A \rho)$ depends on the system state $\rho$, which includes a factor $z$. Specifically, $z$ is defined as the solution of the eigenvalue equation $a\,|z\rangle = z\,|z\rangle$, where $a$ is the annihilation operator. 
In strongly dissipative regimes, $z$ typically decays exponentially over time~\cite{Cochrane_1999}, so $\operatorname{Tr}(A\rho) $ becomes very small at late times. This implies that the step size constraint on $\Delta t$ becomes increasingly relaxed as the system evolves, especially in the long-time limit.
A detailed analysis of $\operatorname{Tr}(A\rho)$ is provided in the Supplementary Materials~\ref{append:tr_arho}.\\

\textbf{Spatial vs.\ Temporal cost:}
We treat space and time separately: (i) the \emph{spatial} cost is the circuit depth per step to realize the post-selected branch; (ii) {the \emph{temporal} cost is the number of steps needed to reach a physical time $T$ under the step-size restriction above}.
Throughout this section, $d$ denotes the spatial dimension, $L$ the number of grid points, $\mathcal N_R(i)$ the Manhattan neighborhood of radius $R$ at grid point $i$, $K$ the highest spatial-derivative order encoded by the stencil (typically $R=\Theta(K)$, as proved in the Supplementary Materials~\ref{append:rk}), and $r$ the degree of nonlinearity in normal-ordered monomials.

\begin{theorem}[Kraus Rank and Circuit Depth for Linear  Generators]
\label{thm:kraus-spatial-linear}
Consider the linear PDEs of $L$ grid points:
\begin{equation}
\frac{\partial u_i}{\partial t}
\;=\;
\sum_{j\in\mathcal N_R(i)} c_{ij}\,u_j,
\qquad i=1,\ldots,L .
\label{eq:linear-pde}
\end{equation}
The generator corresponding to this system is a linear and positive operator of the form $A=\sum_{(i,j)\in E}\alpha_{ij}\,a_i^\dagger a_j$, where $E=\{(i,j): j\in\mathcal N_R(i)\}$ is a bounded-degree directed stencil and the coefficients $\alpha_{ij}$ are chosen to have the same sparsity as $c_{ij}$. A short-time CPTP realization exists that assigns two non-identity Kraus operators to each directed coupling $(i,j)\in E$. The Kraus rank is therefore $N=2|E|$. For a nearest-neighbor lattice where $|E| = \mathcal{O}(dL)$, the rank becomes $N =\mathcal{O}(dL)$. 
The per-step circuit depth for a binary-tree implementation of this channel scales as $\mathcal{O}(\log N) = \mathcal{O}(\log(dL))$, 
up to additive constants depending on the dimension $d$.
\end{theorem}
Based on Theorem~\ref{thm:kraus-spatial-linear}, we now consider nonlinear partial differential equations whose nonlinearities are polynomial:
\begin{equation}
\frac{\partial u_i}{\partial t}
\;=\;
\sum_{\alpha} c_{i,\alpha} 
\prod_{m=1}^{r} u_{j_m^{(\alpha)}},\  j_m^{(\alpha)}\in\mathcal N_R(i),\ i=1,\ldots,L .
\label{eq:nonlinear-pde}
\end{equation}
The associated \emph{ polynomial} generator uses the same normal-ordered  monomials and support, is defined as
$A=\sum_{i,\alpha} L_{i,\alpha}$ with $
L_{i,\alpha}=a_i^\dagger \prod_{m=1}^{r} a_{j_m^{(\alpha)}}$.
Here $r$ is the degree of nonlinearity and the factors $a_{j_m^{(\alpha)}}$ act within the neighborhood $\mathcal N_R(i)$.

\begin{theorem}[Kraus Rank and Circuit Depth for Polynomial  Generators]
\label{thm:rank-polynomial}
On a $d$-dimensional lattice with $L$ grid points, nonlinearity degree $r$, and stencil order $K$, the total Kraus rank is $\mathcal{O} \big(L\,K^{dr}\big)$ and the depth of the binary-tree per-step is $\mathcal O(\log N)=\mathcal O \big(\log L + dr\log K\big)$.
\end{theorem}
The proofs of Theorems~\ref{thm:kraus-spatial-linear}–\ref{thm:rank-polynomial} are provided in the Supplementary Materials~\ref{appen:overhead-proof2}.\\

From Eq.~\eqref{eq:stepsize-constraint}, the number of successful steps to reach \(T\) is at least $\Theta\big(T\Tr(A\rho)/\epsilon\big)$. For fixed \(K,r\), each physical step costs \(\mathcal O(\log L+ dr\log K)\) depth, while the step count is controlled by \(\Tr(A\rho)\).  The total cost to reach time \(T\) is therefore upper-bounded by
\[
\underbrace{\mathcal O(\log L+ dr\log K)}_{\text{depth per step}}
\times
\underbrace{\Theta\big(T\Tr(A\rho)/\epsilon\big)}_{\text{accepted steps}}.
\]
In practice, the actual Kraus rank, and the total cost can be much less than the given bounds.

\section{Numerical Simulation}
\label{sec:cohe-mc}
We simulate our continuous-variable quantum algorithm in a coherent-state framework. Rather than evolving the full density operator, we propagate observables by tracking coherent amplitudes and their variances. In this section, we focus on 1D Burgers' equation and 2D Fisher-KPP dynamics. At each grid point we assume a Gaussian readout model whose mean equals to the deterministic PDE solution and whose variance is predicted by the first-order moment formulas derived in Supplementary Materials~\ref{app:1D} and \ref{app:2D}.  At each saved time we sample $N = 10^4$ i.i.d. shots per grid point, compute the sample mean and the unbiased sample variance \(\Var_{\rm em}\), and compare them with the theoretical predictions $u$ and $\Var_{\mathrm{th}}$; the expected relative $1\sigma$ fluctuation of $\Var_{\mathrm{em}}$ is $\sqrt{2/(N-1)}$. Fig.~\ref{fig:burgers-mc-validation} and Fig.~\ref{fig:fisher2d-mc-validation} present the corresponding mean field and the sample-mean residual, while Table~\ref{tab:burgers-var} and Table~\ref{tab:fisher2d-var} compare $\Var_{\mathrm{em}}$ and $\Var_{\mathrm{th}}$.

In the Supplementary Materials we also describe a Trotterization-based algorithm for the lid-driven cavity Navier–Stokes flow and provide the tensor network simulation of the Burgers' equation for demonstrating the Trotterization algorithm's feasibility.

\subsection{1D Burgers’ Equation}
\label{sec:burger1d-main}
\begin{figure*}
  \centering
  \includegraphics[width=.9\textwidth]{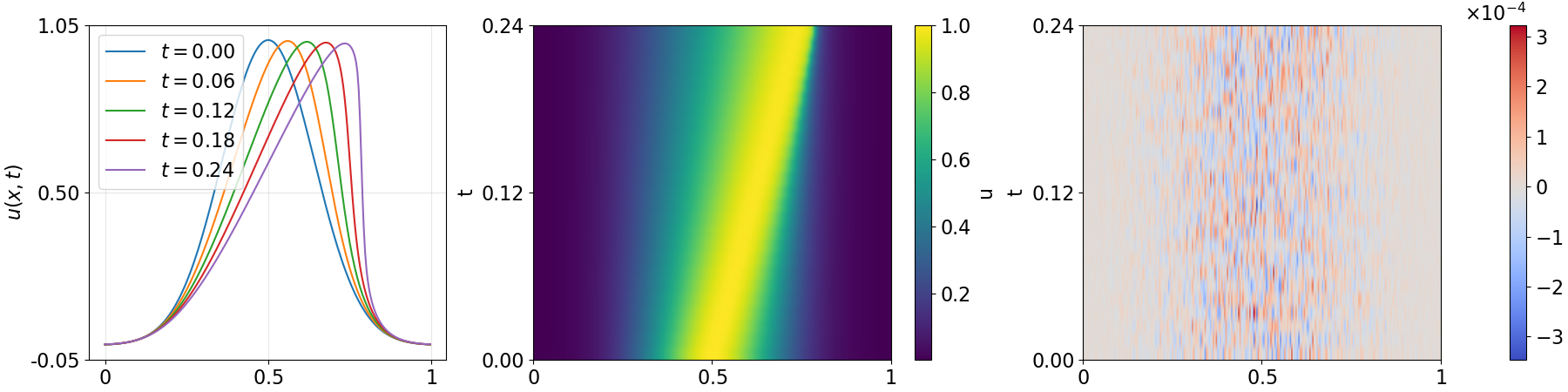}
  \caption{\textbf{1D Burgers' validation.}
  \emph{Left}: mean-field solution profiles at selected times. The results demonstrate the rightward advection with speed set by $u$, nonlinear steepening from the advective term $-u {\partial u}/{\partial x}$ together with diffusive broadening from ${1/R_e}\cdot{\partial^2 u}/{\partial x^2}$. 
  \emph{Middle}: spacetime map of the mean field $u(x,t)$ over the entire evolution window, showing advection-dominated rightward drift with weak viscous spreading.  
  \emph{Right}: Sample-mean bias using $N=10^4$ shots per grid point. The residuals are centered around zero and spatially unstructured, consistent with the prediction that $\Var_j(t)$ remains constant and time-independent to first order [Eq.~\eqref{eq:var-burgers-main}], so that sampling fluctuations are the only visible deviations.
  }
  \label{fig:burgers-mc-validation}
\end{figure*}

The viscous Burgers’ equation is a canonical nonlinear advection–diffusion model and a standard testbed for transport with viscosity:
\begin{equation}
\frac{\partial u}{\partial t}
=
-  u\frac{\partial u}{\partial x} +
\frac{1}{R_e}\frac{\partial^2 u}{\partial x^2}.
\label{eq:burgers-pde}
\end{equation}
In the coherent-state simulation, we propagate a single deterministic mean-field solution \(u(x,t)\) using the Forward-Euler update rule in Eq.~\eqref{eq:burgers-euler}, and adopt a Gaussian readout model with variance given by the first-order moment formula in Eq.~\eqref{eq:burgers-var} at each grid point. For each saved time between 0 and 0.24\,s we draw \(N=10^4\) i.i.d.\ shots per grid point, compute the sample mean and the unbiased sample variance \(\Var_{\rm em}\), and compare them with the theoretical predictions \(u\) and \(\Var_{\rm th}\). The expected relative \(1\sigma\) fluctuation of \(\Var_{\rm em}\) is \(\sqrt{2/(N-1)}\approx 1.414\times10^{-2}\). Table~\ref{tab:burgers-var} reports the spatially averaged theoretical and empirical variances, while Fig.~\ref{fig:burgers-mc-validation} shows the mean fields and the sample-mean residuals. The sample-mean residuals are zero-mean with a speckle-like appearance and nearly time-invariant contrast, consistent with a leading-order time-independent readout variance \(\Var_{\rm th}\) [Eq.~\eqref{eq:burgers-var}]. A tensor network simulation of the Burgers' equation for demonstrating the Trotterization algorithm's feasibility is also provided in the Supplementary Materials \ref{sec:appe}.

\begin{table}[t]
\centering
\footnotesize
\setlength{\tabcolsep}{3.5pt}
\begin{tabular}{c c c c c}
\toprule
$t$ (s) & $\overline{\Var}_{\rm th}$ & $\overline{\Var}_{\rm em}$ & relBias & rel\,$L^2$ \\
\midrule
0.00 & 3.534144 & 3.532065 & $-5.883{\times}10^{-4}$ & $1.398{\times}10^{-2}$ \\
0.06 & 3.534144 & 3.533156 & $-2.795{\times}10^{-4}$ & $1.523{\times}10^{-2}$ \\
0.12 & 3.534144 & 3.532177 & $-5.567{\times}10^{-4}$ & $1.529{\times}10^{-2}$ \\
0.18 & 3.534144 & 3.530277 & $-1.094{\times}10^{-3}$ & $1.546{\times}10^{-2}$ \\
0.24 & 3.534144 & 3.525037 & $-2.577{\times}10^{-3}$ & $1.380{\times}10^{-2}$ \\
\bottomrule
\end{tabular}
\caption{\textbf{Burgers variance (empirical vs.\ theoretical).}
All reported variances are of order $10^{-5}$ or smaller, underscoring that noise around the mean-field trajectory remains extremely weak.
We report $\overline{\Var}_{\rm th}$, $\overline{\Var}_{\rm em}$, the relative bias $\mathrm{relBias}=(\overline{\Var}_{\rm em}-\overline{\Var}_{\rm th})/\overline{\Var}_{\rm th}$, and the per-time global relative $L^2$ error $\mathrm{rel}\ L^2=\|\Var_{\rm em}-\Var_{\rm th}\|_2/\|\Var_{\rm th}\|_2$; an overbar denotes an average over grid points.
For $N=10^4$, the theoretical sampling envelope is $\sqrt{2/(N-1)}=1.414{\times}10^{-2}$, which lies within the per-time relative $L^2$ values reported in the table and is consistent with finite-$N$ sampling noise.}
\label{tab:burgers-var}
\end{table}

\subsection{2D Fisher-KPP Dynamics}
\label{sec:fisher2d-main}
\begin{figure*}
  \centering
  \includegraphics[width=.98\textwidth]{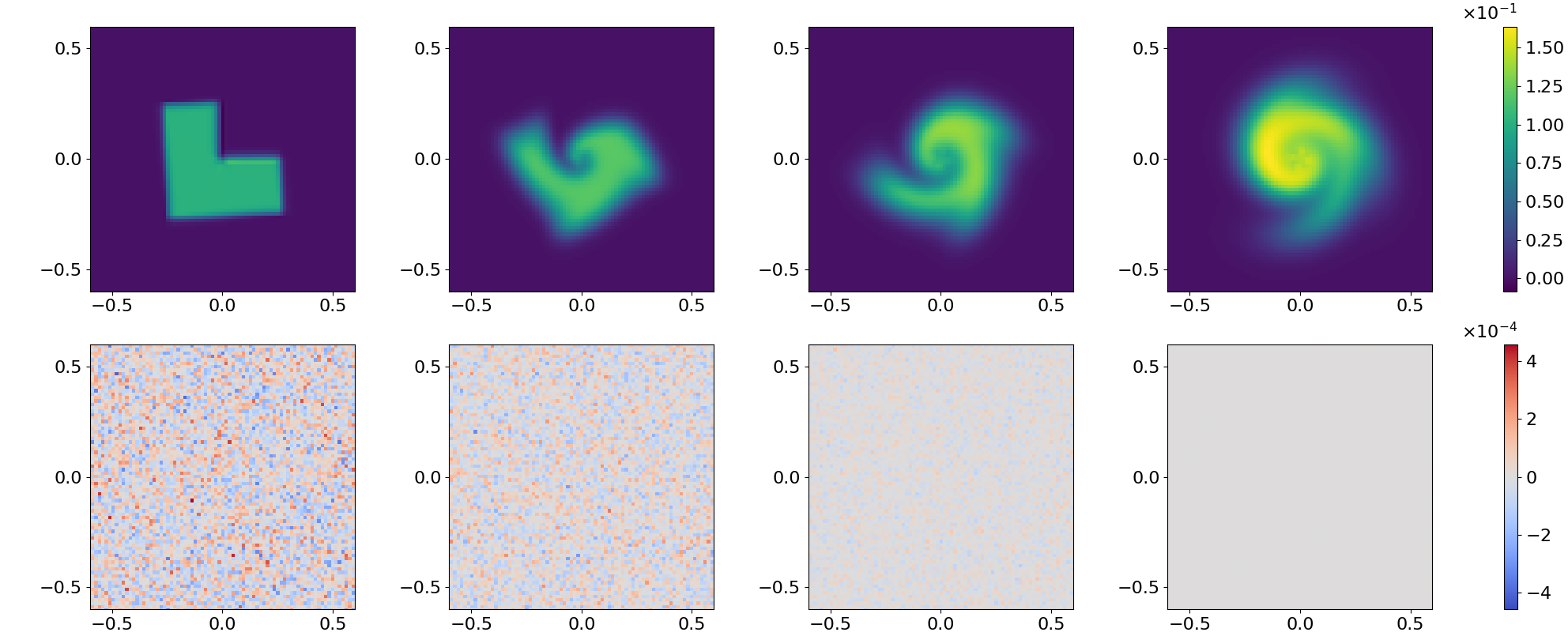}
  \caption{\textbf{2D Fisher-KPP validation.}
  Row\,1: analytic mean fields $u(x,y,t)$ at representative times $t=0.02,\,0.20,\,0.40,\,0.80$\,s, showing the rotational advection imposed by the velocity field, the smoothing action of diffusion ($\mathrm{Pe}=200$), and the amplitude saturation from the Fisher reaction term ($\mathrm{Da}=1$).  
  Row\,2: Bias maps from $N=10^4$ trajectories.  The residuals remain centered around zero with no coherent structure, confirming that the stochastic sampling is unbiased.  
  The omitted panels for the predicted $1\sigma$ noise width and the empirical-to-analytic error-bar ratio behave as in Supplementary Materials~\ref{app:2D-fisher}, i.e.\ Gaussian fluctuations with variance $V_{p,q}(t)$ that scale as $\sigma/\sqrt{N}$.  Together, the results validate the analytic update formulas \eqref{eq:fisher2d-mean}--\eqref{eq:fisher2d-var} across the evolving vortex and front structures.}
  \label{fig:fisher2d-mc-validation}
\end{figure*}

We next consider the 2D Fisher–KPP equation as a canonical reaction–advection–diffusion benchmark to illustrate our continuous-variable framework on a rotating flow.
The 2D Fisher-KPP equation involves the diffusion and reaction of a scalar variable advected by a known velocity field.  
\begin{equation*}
    \frac{\partial u}{\partial t} + v_x\frac{\partial u}{\partial x} + v_y \frac{\partial u}{\partial y} = \frac{1}{\mathrm{Pe}}\left(\frac{\partial^2 u}{\partial x^2} + \frac{\partial^2 u}{\partial y^2}\right) + \mathrm{Da}\left[u(1-u)\right].
\end{equation*}
The derivations of the mean and variance of the corresponding $z$ are provided in the Supplementary Materials~\ref{app:2D-fisher}.
We quantitatively compare the theoretical variance $\Var_{\rm th}$ from Eq.~\eqref{eq:fisher2d-var} with the empirical variance $\Var_{\rm em}$ obtained from $N=10^4$ shots per grid point in Table~\ref{tab:fisher2d-var} (theoretical relative $1\sigma$ for the sample variance is $\sqrt{2/(N-1)}\approx1.414\times10^{-2}$).  Spatial means and global relative bias are reported in Fig.~\ref{fig:fisher2d-mc-validation};  results agree with theory within sampling fluctuations across the evolving vortex structures. This result confirms that Eqs.~\eqref{eq:fisher2d-mean}–\eqref{eq:fisher2d-var} correctly capture rotational transport, diffusion, and nonlinear growth to first order in $\Delta t$.

\begin{table}[t]
\centering
\footnotesize
\setlength{\tabcolsep}{3.5pt}
\begin{tabular}{c c c c c}
\toprule
$t$ (s) & $\overline{\Var}_{\rm th}$ & $\overline{\Var}_{\rm em}^{\rm mean}$ & relBias & rel\,$L^2$ \\
\midrule
0.02 & 14.067745 & 14.067418 & $-2.326{\times}10^{-5}$ & $+1.414{\times}10^{-2}$ \\
0.20 &  6.326569 &  6.325933 & $-1.005{\times}10^{-4}$ & $+1.407{\times}10^{-2}$ \\
0.40 &  0.997785 &  0.997726 & $-5.907{\times}10^{-5}$ & $+1.410{\times}10^{-2}$ \\
0.80 &  0.001421 &  0.001421 & $+2.130{\times}10^{-4}$ & $+1.402{\times}10^{-2}$ \\
\bottomrule
\end{tabular}
\caption{\textbf{Fisher-KPP variance (empirical vs.\ theoretical).}
Reported variances are $\mathcal{O}(10^{-5})$, consistent with extremely weak noise.
We report $\overline{\Var}_{\rm th}$, $\overline{\Var}_{\rm em}$, the relative bias $\mathrm{relBias}=(\overline{\Var}_{\rm em}-\overline{\Var}_{\rm th})/\overline{\Var}_{\rm th}$, and the per-time global relative $L^2$ error $\mathrm{rel}\ L^2=\|\Var_{\rm em}-\Var_{\rm th}\|_2/\|\Var_{\rm th}\|_2$; the bar denotes averaging over grid points. 
For $N=10^4$, the theoretical sampling envelope is $\sqrt{2/(N-1)}=1.414{\times}10^{-2}$, which lies within the per-time relative $L^2$ values reported here.}
\label{tab:fisher2d-var}
\end{table}

\section{Noise Resilience}\label{sec:noise}

The influence of environmental noise on the bosonic quantum simulation is captured by open-system dynamics. The evolution of the density matrix $\rho$ of the mixed bosonic state is governed by a Lindblad master equation that incorporates both the coherent generator $M$ and the dominant noise channel. Photon loss is considered as the primary decoherence mechanism in typical bosonic platforms \cite{knill2001scheme,leviant2022quantum,harris2025logical}. The master equation is written as:
\begin{equation} \label{eq:mastereq}
    \frac{\partial \rho}{\partial t} = M\rho + \rho M^\dagger
    + \sum_{j=1}^n\left[\gamma a_j \rho a_j^\dagger - \frac{\gamma}{2}\{a_j^\dagger a_j, \rho\}\right],
\end{equation}
where $\gamma$ denotes the photon loss rate. The dissipator corresponds to amplitude damping. For analytical tractability, the state is expressed in the diagonal coherent-state representation via
\begin{equation}
    \rho(t) = \int P(z,z^*,t) |z\rangle \langle z| \, d^2z,
\end{equation}
with $d^2z = \prod_{j=1}^n d^2 z_j$ and $P(z,z^*,t)$ being a quasi-probability distribution (the generalized Glauber-Sudarshan $P$-function) \cite{breuer}. Substitution of this ansatz into the master equation leads to a partial differential equation for $P$, which, after standard manipulations, takes the form of a generalized Fokker–Planck equation~\eqref{appB:PDE}.
Under the assumption that nonlinearities in $F_j$ are sufficiently smooth and 
% {\color{red}To Hirad: How can we justify this approximation?}
that $P$ remains well approximated by a Gaussian-like peak in phase space, the evolution of the first moments (the coherent amplitudes) is extracted.  The drift terms emerging from this analysis yield the following effective evolution for the coherent amplitudes:

\begin{equation} \label{eq:coherent_dynamics_noise}
    \frac{d z_k}{dt} = F_k(z) - \frac{\gamma}{2} z_k
\end{equation}
where the $-\frac{\gamma}{2}z_k$ term represents exponential decay due to photon loss. The details of this derivation are provided in the Supplementary Materials \ref{sec:appb}. The dominant degradation is therefore a multiplicative decay of amplitude with rate $\gamma/2$. When higher-order dissipative effects become non-negligible, corrections proportional to $\gamma^2$ or nonlinear couplings between loss and the dynamics can be derived by retaining higher cumulants in the expansion of $P$. 
In parallel, stochastic unravelings such as the quantum-trajectory (Monte Carlo wavefunction) method provide an approximation of the Lindblad dynamics for any $\gamma$.  Therefore, they can be used to benchmark and quantify deviations from the truncated Gaussian-peak approximation \cite{breuer,Daley2014trajectories}.\\

The solution of  Eq.~\eqref{eq:coherent_dynamics_noise} is
\begin{equation} \label{eq:errormitig}
    z_k(t) = \tilde{z}_k(t) \exp\left(-\frac{\gamma}{2} t\right),
\end{equation}
where $\tilde{z}_k(t)$ denotes the ideal noiseless evolution satisfying $\frac{d\tilde{z}_k}{dt} = F_k(\tilde{z})$. 
Consequently, the measured expectation values are attenuated relative to their ideal counterparts, and this systematic decay can be compensated for if the loss rate is estimated.
Therefore, a time-local correction (counterterm) is implemented by multiplying the solution profile by an exponential term $\exp\left(\bar{\gamma} t/2\right)$ where $\bar{\gamma}$ is the estimate of $\gamma$. This term effectively inverts exponential damping. The estimated loss rate $\bar{\gamma}$ can be characterized experimentally through coherent state decay measurements and performed on a calibration schedule to infer the effective Lindblad parameters.\\

Residual errors arising from imperfect knowledge of $\gamma$, higher-order cross terms, or non-Gaussian distortions can be further suppressed by combining the analytic correction with extrapolation techniques. For example, a sequence of short-time evolutions with varying effective loss rates (achieved by artificially inserting calibrated attenuation) can be used to perform Richardson extrapolation back to the zero-loss limit \cite{temme2017error,mele2024quantum}. In addition, measurement-based feedback can be used: after each time step, the coherent-state amplitude is estimated, the discrepancy from the expected corrected value is evaluated, and a small coherent displacement is applied to the recenter state, effectively performing a form of state stabilization \cite{wiseman1993quantum}.\\

The validity of the Gaussian ansatz and linear correction is guaranteed when the loss rate remains small compared to the inverse time step ($\Delta t$) and when the nonlinear vector field $F(z)$ does not drive the system into regions of phase space where higher moments become dominant. In stiff or strongly nonlinear regimes, adaptive step-sizing or higher-order noise-aware integrators can be used, with noise-modified commutator expansions in the Trotterization algorithm guiding the required modifications (introduced in Supplementary Materials~\ref{appsec:lid driven}).

\section{Conclusion}\label{sec:conc}

We  have introduced a provably efficient, continuous-variable quantum algorithm for solving nonlinear PDEs. The key innovation of our work is the application of the KvN formalism to lift nonlinear dynamics into a linear, albeit dissipative, evolution within an infinite-dimensional bosonic Hilbert space.\\

This method allows for the encoding of classical fields into coherent states of bosonic modes. The dynamics is then implemented as a CPTP map, which we compile into a quantum circuit with a logarithmic depth per time step. Specifically, for a $d$-dimensional system with $L$ lattice grid points, a polynomial nonlinearity of degree $r$, and spatial derivatives of order $K$, the {\color{red}} per-step circuit depth scales as $O(\log L + dr \log K)$.\\

This continuous-variable framework provides a significant advantage by sidestepping the truncation errors and substantial resource overhead associated with Carleman linearization-based digital methods. Our high-fidelity numerical simulations of Fisher-KPP equations and the lid-driven cavity problem serve as a robust validation of the algorithm's accuracy and near-term feasibility. Furthermore, we argue that certain forms of noise resilience will occur for highly dissipative quantum systems. The KvN formalism, traditionally a theoretical tool for unifying classical and quantum mechanics, is shown here to be a powerful computational framework, analogous to a kernel method for dynamical systems. By physically realizing the enlarged linear space of the KvN formalism, our approach offers a direct and efficient means of simulating complex nonlinear phenomena. This work establishes a compelling alternative to fault-tolerant digital quantum computing, paving a viable path toward achieving a potential quantum speedup in scientific computing on near-term analog hardware. Future research will be directed towards the experimental implementation of this algorithm and the creation of higher-order, more noise-resilient integrators.\\

We note that the famous connection between the KvN approach and Carleman linearization has been widely noted in the quantum algorithm community \cite{joseph2020koopman,lin2022koopman,novikau2025quantum,higuchi2025quantum,ito2023map,luo2023quack,tanaka2023polynomial}. Compared to those works, our primary contribution is to connect KvN with quantum channel and Kraus operator frameworks in quantum information science, and to provide solid arguments about the scaling statement in quantum algorithms. From the digital computing point of view, our algorithm takes a significant leap from high fault-tolerant overhead towards more near-term-feasible routines. A related line of research is the Schr\"{o}dingerisation method~\cite{jin2024quantum,golse2025quantum,Jin_2024112707,Jin_2024}, which maps linear or nonlinear PDEs to a Schr\"{o}dinger-type equation solvable by Hamiltonian simulation after spatial discretization. Although based on the digital truncation approach instead of continuous-variable quantum computing, their approach can, in principle, be combined with the KvN formalism we have developed and runnable on analog quantum hardware. \\

\textit{Acknowledgments}---We thank Ken Brown, Andrew Childs, Fred Chong, Jens Eisert, Steve Girvin, Hsin-Yuan Huang, Travis Humble, Liang Jiang, Jin-Peng Liu, and John Preskill for helpful discussions. The authors acknowledge support from the U.S.\ Air Force Office of Scientific Research (AFOSR) under Grant No. FA9550-23-1-0014. YG, JL, JC, ZW and GL are supported in part by the University of Pittsburgh School of Computing and Information, Department of Computer Science, Pitt Cyber, PQI Community Collaboration Awards, John C. Mascaro Faculty Scholar in Sustainability, NASA under award number 80NSSC25M7057, and Fluor Marine Propulsion LLC (U.S. Naval Nuclear Laboratory) under award number 140449-R08. This work has been co-authored by a contractor of the U.S. Government under contract number DOE89233018CNR000004. Accordingly, the U.S. Government retains a non-exclusive, royalty-free license to publish or reproduce the published form of this contribution, or allow others to do so, for U.S. Government purposes.

\bibliography{main}
\onecolumngrid
\newpage

\vspace{0.5in}

\begin{center}
	{\Large \bf Supplementary Materials}
\end{center}

\let\addcontentsline\oldaddcontentsline
\tableofcontents
\appendix

\section{Technical derivations for the KvN formalism}
\label{app:tech-derive-kvn}
\subsection{Derivation of coherent-state operator identities}
\label{subsection:deri-coherets}

In this subsection we prove Segal–Bargmann identities and for the purpose of the derivation, we adopt the unnormalized coherent state 
\begin{equation}
  \ket{z} =  e^{z\,a^\dagger}\ket{0}, \qquad z\in\mathbb{C}.
  \label{eq:coh-def}
\end{equation}

We will repeatedly use the power-series expansion of the exponential and the definition/action of the Fock basis:
\begin{equation}
  e^{z\,a^\dagger}=\sum_{n=0}^{\infty}\frac{z^n}{n!}\,(a^\dagger)^n,
  \label{eq:exp-series}
\end{equation}
\begin{equation}
  \ket{n}=\frac{(a^\dagger)^n}{\sqrt{n!}}\ket{0}, 
  \qquad 
  a\ket{n}=\sqrt{n}\,\ket{n-1}.
  \label{eq:fock-def-action}
\end{equation}
Combining Eqs.~\eqref{eq:coh-def}--\eqref{eq:fock-def-action}, we can get the number-basis expansion of the coherent state:
\begin{equation}
  \ket{z} = \sum_{n=0}^{\infty}\frac{z^n}{n!}\,(a^\dagger)^n \ket{0} = \sum_{n=0}^{\infty} \frac{z^n}{\sqrt{n!}}\frac{(a^\dagger)^n}{\sqrt{n!}}\,\ket{0}= \sum_{n=0}^{\infty}\frac{z^n}{\sqrt{n!}}\,\ket{n}.
  \label{eq:coh-number-exp}
\end{equation}

Acting with $a$ on Eq.~\eqref{eq:coh-number-exp} and using $a\ket{n}=\sqrt{n}\ket{n-1}$ from Eq.~\eqref{eq:fock-def-action}, we obtain
\begin{equation}
  a\ket{z} 
    = \sum_{n=0}^{\infty}\frac{z^n}{\sqrt{n!}}\,a\ket{n}
    = \sum_{n=1}^{\infty}\frac{z^n}{\sqrt{n!}}\,\sqrt{n}\,\ket{n-1} \notag\\
    = \sum_{m=0}^{\infty}\frac{z^{m+1}}{\sqrt{m!}}\,\ket{m}
    = z \sum_{m=0}^{\infty}\frac{z^{m}}{\sqrt{m!}}\,\ket{m}
    = z\,\ket{z},
  \label{eq:a-on-coh}
\end{equation}
which proves the first identity in Segal–Bargmann identities.

For the derivation of second equation in  Segal–Bargmann identities, we can differentiate the definition \eqref{eq:coh-def} term-by-term, it can yield:
\begin{equation}
    \frac{\partial}{\partial z}\ket{z} = \sum_{n=1}^{\infty}\frac{n z^{n-1}}{n!}\,(a^\dagger)^n \ket{0} = \sum_{m=0}^{\infty}\frac{z^m}{m!}\,(a^\dagger)^{m+1} \ket{0} = a^{\dagger} \ket{z}
\end{equation}

\subsection{Recovering the classical ODE from the KvN lift}
\label{subsection:reco-clsskvn}

We assume the multivariable KvN setup in the main text:
\begin{equation*}
\frac{d}{dt} z_j(t)=F_j \big(z(t),t\big),\qquad
A(t)=\sum_{j=1}^{n} a_j^\dagger F_j(a,t),
\end{equation*}
with the coherent-state identities
$a_k\ket{z}=z_k\ket{z}$,
$F_j(a,t)\ket{z}=F_j(z,t)\ket{z}$,
and $\bra{z}a_j^\dagger=z_j^*\bra{z}$
for the unnormalized product coherent state
$\ket{z}=\exp(\sum_\ell z_\ell a_\ell^\dagger)\ket{0}$.
Define
\begin{equation*}
N_k(t):=\bra{ z(t)}a_k\ket{z(t)},\ 
Z(t):=\bra{z(t)}\ket{z(t)},\ 
z_k(t)=\frac{\bra{ z(t)}a_k\ket{z(t)}}{Z(t):=\bra{z(t)}\ket{z(t)}} =\frac{N_k(t)}{Z(t)}.
\end{equation*}
Differentiating the quotient term $z_k(t)$ gives
\begin{equation*}
\dot z_k
=\frac{\dot N_k\,Z-N_k\,\dot Z}{Z^2}.    
\end{equation*}
Using the lifted evolution from Eq.~\eqref{eq: lifted evolution} and the conjugate transpose version
$\bra{\dot z}= \bra{z}A^\dagger$, we get
\begin{equation*}
    \dot N_k
=\bra{\dot z}a_k\ket{z}+\bra{ z}a_k\ket{\dot z}
= \bra{z}A^\dagger a_k\ket{z}+\bra{ z}a_kA\ket{z},
\ 
\dot Z=\bra {z}(A+A^\dagger)\ket{z}.
\end{equation*}
We now evaluate the matrix elements.
\begin{itemize}
    \item[(i)] For the term $\bra{ z}a_kA\ket{z}$, using $[a_k,a_j^\dagger]=a_ka_j^{\dagger}-a_j^{\dagger}a_k=\delta_{kj}$ we can have
        \[
    \begin{aligned}
    \bra{z}a_kA\ket{z}
    &=\sum_j \bra{z}a_k a_j^\dagger F_j(a,t)\ket{z}
    =\sum_j \bra{ z}\left(a_j^\dagger a_k+\delta_{kj}\right)F_j(a,t)\ket{z} \\
    &=\sum_j \left(z_j^* z_k\,F_j(z,t)+\delta_{kj} F_j(z,t)\right)\,Z
    =\left(z_k\sum_j z_j^*F_j(z,t)+F_k(z,t)\right)Z.
    \end{aligned}
    \]

    \item[(ii)] For $\langle z|A^\dagger a_k|z\rangle$, writing $A^\dagger=\sum_j F_j(a,t)^\dagger a_j$ and noting that $F_j(a,t)^\dagger$ is a polynomial in creation operators, normal ordering gives
      \[
      \langle z|F_j(a,t)^\dagger=F_j(z,t)^*\,\langle z|,
      \]
      hence
      \[
      \langle z|A^\dagger a_k|z\rangle
      = \sum_j \langle z|F_j(a,t)^\dagger a_j a_k|z\rangle
      =\sum_j F_j(z,t)^*\langle z|a_j a_k|z\rangle
      =z_k\sum_j z_j F_j(z,t)^*\,Z.
      \]

    \item[(iii)] As for the normalization term, we have
        \[
    \langle z|(A+A^\dagger)|z\rangle
    =\sum_j \big(z_j^*F_j(z,t)+z_j F_j(z,t)^*\big)\,Z
    =2\,\mathrm{Re} \sum_j z_j^*F_j(z,t)\,Z.
    \]
\end{itemize}
Collecting (i)–(iii), we have
\[
\begin{aligned}
\dot z_k
&=\frac{1}{Z}\Big[\langle z|a_kA|z\rangle+\langle z|A^\dagger a_k|z\rangle\Big]
-\frac{N_k}{Z^2}\,\langle z|(A{+}A^\dagger)|z\rangle \\
&=\Big(z_k\sum_j z_j^*F_j+F_k\Big)
+\Big(z_k\sum_j z_j F_j^*\Big)
-z_k\cdot 2\,\mathrm{Re} \sum_j z_j^*F_j \\
&=F_k(z(t),t).
\end{aligned}
\]
The terms proportional to $z_k$ cancel exactly, leaving
$\dot z_k(t)=F_k(z(t),t)$, which recovers the original ODE.
\hfill$\square$

\section{Quantum Channel Execution and Sampling Overhead}\label{appen:overhead-proof2}

\subsection{$\operatorname{Tr}(A \rho)$ Calculation}
\label{append:tr_arho}
We aim to compute the value $\operatorname{Tr}(A \rho)$, where $A$ is the system generator. The derivation proceeds as follows:
\begin{equation*}
\begin{aligned}
\operatorname{Tr}(A \rho) 
&= \operatorname{Tr}\big(A \ketbra{z}{z}\big) = \bra{z}A\ket{z} \\
&= \bra{z} \frac{d}{dt} \ket{z} = \sum_j \bra{z} a_j^\dagger F_j(z,t) \ket{z} \\
&= \sum_j \bra{z} a_j^\dagger \ket{z} \cdot F_j(z,t) \\
&= \sum_j z_j^* \cdot F_j(z,t),
\end{aligned}
\end{equation*}
where the last step utilizes the well-known property of coherent states \(\bra{z} a_j^\dagger \ket{z} = z_j^*\), with \( z_j^* \) denoting the complex conjugate of \( z_j \). This result follows from the fact that a coherent state \( |z\rangle \) is an eigenstate of the annihilation operator \(a_j\), so the expectation value of the creation operator \(a_j^\dagger\) is the complex conjugate of the eigenvalue:
\(
a_j |z\rangle = z_j |z\rangle \Rightarrow \bra{z} a_j^\dagger |z\rangle = z_j^*.
\)
Therefore, once the functions \( z_j(t) \) are determined, they can be substituted into \( F_j(z,t) \) to evaluate \( \operatorname{Tr}(A \rho) \).

{
\color{black}
\subsection{Justification of the Stencil Radius Scaling: \texorpdfstring{$R = \Theta(K)$}{R = Theta(K)}}\label{append:rk}

Here, we justify the scaling relation $R = \Theta(K)$ for a compact finite-difference stencil on a uniform grid, where $R$ is the Manhattan radius of the stencil and $K$ is the order of the spatial derivative to be approximated. We establish this by proving a necessary lower bound and a constructive upper bound on $R$.

\paragraph{Necessity (Lower Bound).}
To show that $R$ must grow at least linearly with $K$, we analyze the constraints required to construct a derivative stencil. Consider the simplest case: a one-dimensional, symmetric, centered stencil of radius $R$ used to approximate the $K$-th derivative, $f^{(K)}(x)$. The approximation has the form:
\begin{equation}
    D^{(K)}f(x) := \frac{1}{h^K} \sum_{m=-R}^{R} c_m f(x+mh) \approx f^{(K)}(x),
\end{equation}
where $h$ is the grid spacing and $\{c_m\}$ are the stencil coefficients, with $c_{-m} = (-1)^K c_m$ for symmetry. To ensure accuracy, we expand $f(x+mh)$ in a Taylor series around $x$:
\begin{equation}
    D^{(K)}f(x) = \frac{1}{h^K} \sum_{m=-R}^{R} c_m \sum_{j=0}^{\infty} \frac{(mh)^j}{j!} f^{(j)}(x) = \sum_{j=0}^{\infty} \left( \frac{1}{j!} \sum_{m=-R}^{R} m^j c_m \right) f^{(j)}(x).
\end{equation}
For this expression to approximate $f^{(K)}(x)$, the coefficients of the Taylor series must match. Specifically, the stencil must annihilate all lower-order derivative terms and correctly reproduce the $K$-th order term:
\begin{enumerate}
    \item $\sum_{m=-R}^{R} m^j c_m = 0$ for all $j < K$.
    \item $\sum_{m=-R}^{R} m^K c_m = K!$.
\end{enumerate}
Due to the symmetry, conditions for $j$ with parity different from $K$ are automatically satisfied. This leaves approximately $K/2$ linear conditions from the first requirement, plus one condition from the second, for a total of $\lceil K/2 \rceil + 1$ constraints. The number of independent variables we can choose is the number of unique coefficients, which is $R+1$ (i.e., $\{c_0, c_1, \dots, c_R\}$). For a solution to exist, the number of variables must be at least the number of constraints:
\begin{equation}
    R+1 \geq \lceil K/2 \rceil + 1 \implies R \geq \lceil K/2 \rceil.
\end{equation}
This establishes the necessary lower bound $R = \Omega(K)$. This one-dimensional logic extends to higher dimensions by considering the stencil's restriction to any coordinate axis.

\paragraph{Sufficiency (Upper Bound).}
To show that $R$ need not grow faster than linearly with $K$, we demonstrate that a solution can always be constructed. The system of linear equations for the coefficients $\{c_m\}$ derived above is known to have a unique solution for $R \geq \lceil K/2 \rceil$. For any pure $K$-th derivative in one dimension, an explicit formula with the minimal radius $R = \lceil K/2 \rceil$ can be found.

For mixed derivatives in $d>2$ dimensions, such as $\partial^{p_1+\dots+p_d}f / (\partial x_1^{p_1} \dots \partial x_d^{p_d})$ where $\sum p_i = K$, we can construct the full stencil by composing 1D stencils along each axis. The 1D stencil for the $p_i$-th derivative has a radius of $\lceil p_i/2 \rceil$. The total Manhattan radius of the composite stencil is the sum of the individual radii:
\begin{equation}
    R = \sum_{i=1}^{d} \lceil p_i/2 \rceil \leq \sum_{i=1}^{d} (p_i/2 + 1) = \frac{K}{2} + d.
\end{equation}
For fixed dimension $d$, this shows $R$ is bounded by a linear function of $K$, establishing the upper bound $R = \mathcal{O}(K)$.

\paragraph{Conclusion and Example.}
Combining the lower bound $R = \Omega(K)$ and the upper bound $R = \mathcal{O}(K)$, we conclude that the stencil radius must scale linearly with the derivative order: $R = \Theta(K)$.

\textbf{Worked Example ($K=4$ in 1D):} To approximate $f^{(4)}(x)$, the minimal radius is $R = \lceil 4/2 \rceil = 2$. We seek coefficients for the 5-point stencil $\{-2, -1, 0, 1, 2\}$. The constraints are:
\begin{itemize}
    \item $c_0 + 2c_1 + 2c_2 = 0$ \quad (to kill $f(x)$ term, i.e., $j=0$)
    \item $2c_1 + 8c_2 = 0$ \quad (to kill $f''(x)$ term, i.e., $j=2$)
    \item $2c_1 + 32c_2 = 4! = 24$ \quad (to match $f^{(4)}(x)$ term, i.e., $j=4$)
\end{itemize}
Solving this system yields $c_2 = 1$, $c_1 = -4$, and $c_0 = 6$. This gives the well-known 5-point stencil for the fourth derivative:
\begin{equation}
    f^{(4)}(x) \approx \frac{f(x-2h) - 4f(x-h) + 6f(x) - 4f(x+h) + f(x+2h)}{h^4}.
\end{equation}
This achieves the minimal radius $R=2$ for $K=4$, consistent with $R = K/2$.
\paragraph{Connection to Standard Stencils in 2D and 3D.}
The linear relationship $R = \Theta(K)$ is consistent with stencils commonly used in computational physics. For instance, the advection operator ($K=1$) is discretized with first differences, requiring a nearest-neighbor stencil ($R=1$). The diffusion operator or Laplacian ($K=2$) is often approximated with a 5-point (in 2D) or 7-point (in 3D) stencil, which also only involves nearest neighbors ($R=1$). More complex operators like the bi-Laplacian ($K=4$) require a wider 13-point stencil that includes nodes up to two grid points away, giving a radius of $R=2$. In all these standard cases, the radius $R$ is approximately $K/2$, validating the formal scaling relationship.
}

{\color{black}\subsection{Proof of Theorem~\ref{thm:rank-polynomial}: Scaling of the Kraus Rank}\label{}

This section provides the proof for the scaling of the total Kraus rank, $N$, which determines the per-step circuit depth of the algorithm. The theorem states that for a system with $L$ grid points, a nonlinearity of degree $r$, and a stencil of order $K$, the Kraus rank scales as $N = \mathcal{O}(L K^{dr})$.

\paragraph{Premise: Structure of the Generator.}
The evolution generator $A$ is a sum of local terms over all grid points $i$: $A = \sum_{i,\alpha} L_{i,\alpha}$. As defined in the main text, each term $L_{i,\alpha}$ corresponds to a distinct normal-ordered monomial representing a component of the PDE's dynamics. For a nonlinearity of degree $r$, each monomial has the form:
\begin{equation}
    L_{i,\alpha} = a_i^\dagger \prod_{m=1}^{r} a_{j_m^{(\alpha)}},
\end{equation}
where the annihilation operators $a_{j_m^{(\alpha)}}$ act on grid points within the local neighborhood $N_R(i)$ of grid point $i$. The index $\alpha$ enumerates every unique combination of these $r$ annihilation operators. The proof proceeds by counting the number of these unique monomials.

\paragraph{Step 1: Counting Monomials per grid point.}
For a fixed grid point $i$, we need to determine how many unique monomials $L_{i,\alpha}$ can be constructed. This is equivalent to counting the number of distinct products of $r$ annihilation operators, $\prod_{m=1}^{r} a_{j_m}$, where each index $j_m$ is chosen from the set of neighbors $N_R(i)$.

Let $S$ be the number of selectable neighbors in the stencil (excluding the center grid point $i$), and let $S_{\text{eff}}$ be the effective number of choices, where $S_{\text{eff}} = S+1$ if self-coupling (terms like $a_i^r$) is allowed, and $S_{\text{eff}} = S$ otherwise. Because the annihilation operators are bosonic, they commute ($a_j a_k = a_k a_j$), so the order in which we choose the neighbors does not matter. Furthermore, we are allowed to choose the same neighbor multiple times (e.g., $a_k^2$).

This is a classic combinatorial problem: counting the number of multisets of size $r$ drawn from a set of size $S_{\text{eff}}$. The number of such combinations is given by the multiset coefficient:
\begin{equation}
    \#\{\alpha \text{ at grid point } i\} = \binom{S_{\text{eff}} + r - 1}{r}.
\end{equation}

\paragraph{Step 2: Total Kraus Rank Scaling.}
The total number of distinct local monomials over the entire lattice is the number of grid points, $L$, multiplied by the number of monomials per grid point. By the theorem's assumption, each unique monomial corresponds to a constant number, $c_r = \mathcal{O}(1)$, of non-identity Kraus operators. Therefore, the total Kraus rank $N$ is:
\begin{equation}
    N = c_r \cdot L \cdot \binom{S_{\text{eff}} + r - 1}{r}.
\end{equation}
For the purpose of asymptotic scaling, we are interested in the behavior for large stencils ($S_{\text{eff}} \gg r$). The binomial coefficient can be approximated by a polynomial:
\begin{equation}
    \binom{S_{\text{eff}} + r - 1}{r} = \frac{(S_{\text{eff}} + r - 1)\dots(S_{\text{eff}})}{r!} = \mathcal{O}(S_{\text{eff}}^r).
\end{equation}
Thus, the Kraus rank scales as $N = \mathcal{O}(L \cdot S_{\text{eff}}^r)$.

\paragraph{Step 3: Final Expression in Terms of Physical Parameters.}
To complete the proof, we express $S_{\text{eff}}$ in terms of the physical parameters $d$ (dimension) and $K$ (derivative order). The number of grid points in a stencil of Manhattan radius $R$ in $d$ dimensions scales as $S = \Theta(R^d)$. Since $S_{\text{eff}}$ is dominated by $S$, it follows that $S_{\text{eff}} = \Theta(R^d)$.

From the analysis in Supplementary Materials~\ref{append:rk}, we have the established relationship $R = \Theta(K)$. Substituting this into our expression for $S_{\text{eff}}$ gives:
\begin{equation}
    S_{\text{eff}} = \Theta(K^d).
\end{equation}
Finally, substituting this into our scaling law for $N$ yields the main result:
\begin{equation}
    N = \mathcal{O}(L \cdot S_{\text{eff}}^r) = \mathcal{O}(L \cdot (K^d)^r) = \mathcal{O}(L K^{dr}).
\end{equation}
This confirms the scaling of the Kraus rank as stated in Theorem IV.2. The per-step circuit depth, implemented via a binary-tree architecture, scales logarithmically with the rank, giving $\text{Depth} = \mathcal{O}(\log N) = \mathcal{O}(\log L + dr\log K)$.}

\section{Coherent-State Formalism in One Dimension}
\label{app:singlemode}
For a single bosonic mode with annihilation operator $a$, the coherent state $\ket{z}$ is defined by
\begin{equation}
a\ket{z}=z\,\ket{z},\qquad 
\bra{z}a^\dagger=z^*\,\bra{z},\qquad 
\langle a\rangle=\bra{z}a\ket{z}=z,
\end{equation}
where $z\in\mathbb{C}$ is the coherent amplitude and $z^*$ denotes its complex conjugate.  
In the multimode setting with annihilation operators $\{a_j\}_{j=1}^n$, we define the classical field amplitude at time $t$ by
\begin{equation}
z_j(t):=\Tr[a_j \rho(t)] = \langle a_j\rangle_t,
\end{equation}
and, for any normal-ordered polynomial $G(a^\dagger,a)$, we use the coherent-state rule
\(\bra{z}G(a^\dagger,a)\ket{z}=G(z^*,z)\).

We describe the short-time dynamics by the normal-ordered operator
\begin{equation}
M=\sum_{k=1}^{n}a_k^\dagger F_k(a_1,\dots ,a_n),
\label{eq:normal-ordered-M}
\end{equation}
which generates the raw update
\begin{equation}
\rho_{\mathrm{raw}}(t+\Delta t)=e^{M\Delta t}\,\rho(t)\,e^{M^{\dagger} \Delta t}.
\label{eq:raw-map}
\end{equation}
Since $e^{M\Delta t}$ is not generally an isometry (as $M$ need not be anti-Hermitian), the state must be renormalized:
\begin{equation}
\mathcal N(t+\Delta t)=\Tr[\rho_{\mathrm{raw}}(t+\Delta t)],\qquad
\rho(t+\Delta t)=\frac{\rho_{\mathrm{raw}}(t+\Delta t)}{\mathcal N(t+\Delta t)}.
\label{eq:normalize}
\end{equation}
Equivalently, for any observable $O$,
\begin{equation}
\langle O\rangle_{t+\Delta t}
=\frac{\Tr \big[e^{M^\dagger\Delta t}\,O\,e^{M\Delta t}\,\rho(t)\big]}
{\Tr \big[e^{M^\dagger\Delta t}e^{M\Delta t}\rho(t)\big]}.
\label{eq:ratio}
\end{equation}

\subsection{First-Order Moment Dynamics}
\label{app:1D}
\subsubsection{Mean Update: Forward Euler}
To obtain the first-order expansion of the mean field, let $\ket{z}$ denote a product of coherent states with amplitudes $z_k=\langle a_k\rangle_t$, and define the shorthand
\begin{equation}
\Sigma(z):=\sum_{k=1}^n z_k^* F_k(z).
\label{eq:sigma-def}
\end{equation}
Expanding Eq.~\eqref{eq:ratio} with $O=a_j$ yields
\begin{equation}
\langle a_j\rangle_{t+\Delta t}
=\frac{\bra{z}(1+M^\dagger\Delta t)\,a_j\,(1+M\Delta t)\ket{z}}
{\bra{z}(1+M^\dagger\Delta t)(1+M\Delta t)\ket{z}}
+O(\Delta t^2).
\end{equation}
The relevant coherent-state matrix elements are
\begin{align}
\bra{z}a_j M\ket{z}&=F_j(z)+z_j\Sigma(z),\qquad
\bra{z}M^\dagger a_j\ket{z}=z_j\Sigma(z)^*,\\
\bra{z}M\ket{z}&=\Sigma(z),\qquad
\bra{z}M^\dagger\ket{z}=\Sigma(z)^*,
\label{eq:cohestate-matrx}
\end{align}
so that the denominator evaluates to
\begin{equation}
\begin{aligned}
\mathrm{Den}&= \bra{z}(1+M^\dagger\Delta t +M\Delta t+O(\Delta t^2))\ket{z}\\
&=1+\Delta t(\bra{z}M^{\dagger}\ket{z}+\bra{z}M\ket{z})+O(\Delta t^2)) \\
&=1+\Delta t(\Sigma(z)^{\star}+\Sigma(z))+O(\Delta t^2)) \\
&=1+2\Delta t\,\Re\Sigma(z)+O(\Delta t^2).
\end{aligned}
\end{equation}
Similarly, we expand the numerator to first order:
\begin{equation}
\begin{aligned}
\mathrm{Num}&= \bra{z}(1+M^\dagger\Delta t)\,a_j\,(1+M\Delta t)\ket{z}\\
&=\bra{z}(a_j+a_jM\Delta t + M^{\dagger}a_j \Delta t+O(\Delta t^2))\ket{z} \\
&=\bra{z}a_j\ket{z}+\Delta t(\bra{z}a_jM\ket{z}+\bra{z}M^{\dagger}a_j\ket{z})+O(\Delta t^2).
\end{aligned}
\end{equation}
Substituting the matrix elements from Eq.~\eqref{eq:cohestate-matrx} simplifies the numerator to:
\begin{equation}
\begin{aligned}
\mathrm{Num}&=z_j+\Delta t(F_j(z)+z_j\Sigma(z)+z_j\Sigma(z)^{\star})+O(\Delta t^2)\\
&=z_j+\Delta t F_j(z)+2\Delta tz_j\Re\Sigma(z)+O(\Delta t^2)
\end{aligned}
\end{equation}
With both terms expanded, we can now evaluate the full fraction:
\begin{equation}
\langle a_j\rangle_{t+\Delta t}
= \frac{z_j+\Delta t F_j(z)+2\Delta tz_j\,\Re\Sigma(z)}
{1+2\Delta t\,\Re\Sigma(z)}
+O(\Delta t^2)
\end{equation}
To resolve this, we use the first-order Taylor approximation $\frac{1}{1+x}\approx1-x$ for small value $x=2\Delta t\,\Re\Sigma(z)$. This turns the division into a multiplication:
\begin{equation}
\begin{aligned}
\langle a_j\rangle_{t+\Delta t}
&= \left({z_j+\Delta t F_j(z)+2\Delta tz_j\,\Re\Sigma(z)}\right) \left({1-2\Delta t\,\Re\Sigma(z)}\right)
+O(\Delta t^2) \\
&=z_j+\Delta t F_j(z)+2\Delta tz_j\,\Re\Sigma(z) - z_j\cdot 2\Delta t\,\Re\Sigma(z) + O(\Delta t^2)\\\
&=z_j+\Delta t F_j(z)+ O(\Delta t^2)
\end{aligned}
\end{equation}
Therefore,
\begin{equation}
z_j(t+\Delta t)=z_j(t)+\Delta t\,F_j(z(t))+O(\Delta t^2),\qquad
\text{(forward Euler step)}
\label{eq:euler-step}
\end{equation}
which identifies the coherent amplitude with the classical Euler update rule. By definition,
\begin{equation}
z_j(t+\Delta t)=\Tr[a_j \rho(t+\Delta t)].
\label{eq:aj-identity}
\end{equation}

\subsubsection{Variance Update: First Order}
We next turn to fluctuations around the coherent amplitude.  
Define the annihilation–operator variance at grid point $j$:
\begin{equation}
\Var_j(t):=\Var(a_j)
=\langle a_j^\dagger a_j\rangle_t-\big|\langle a_j\rangle_t\big|^2
=\langle (a_j-z_j(t))^\dagger(a_j-z_j(t))\rangle_t.
\label{eq:vjvariance}
\end{equation}
Using Eq.~\eqref{eq:ratio}, introduce the shorthand
\[
E_1=\langle e^{M^\dagger\Delta t}a_j e^{M\Delta t}\rangle_t,\qquad
E_2=\langle e^{M^\dagger\Delta t}a_j^\dagger a_j e^{M\Delta t}\rangle_t,\qquad
D=\langle e^{M^\dagger\Delta t}e^{M\Delta t}\rangle_t,
\]
so that, consistently with \eqref{eq:vjvariance},
\[
\langle a_j\rangle_{t+\Delta t}=\frac{E_1}{D},\qquad
\langle a_j^\dagger a_j\rangle_{t+\Delta t}=\frac{E_2}{D},\qquad
V_j(t+\Delta t)=\frac{E_2D-|E_1|^2}{D^2}.
\]

Expanding each quantity to first order in $\Delta t$ gives
\begin{align}
E_1
&=\langle e^{M^\dagger\Delta t}a_j e^{M\Delta t}\rangle_t
=\langle (1+\Delta t M^\dagger)a_j(1+\Delta t M)\rangle_t+O(\Delta t^2) \nonumber\\
&=\langle a_j\rangle_t+\Delta t\langle M^\dagger a_j+a_j M\rangle_t+O(\Delta t^2)
= z_j+\Delta t A_1+O(\Delta t^2),
\end{align}
with
\begin{equation}
A_1:=\langle M^\dagger a_j+a_j M\rangle_t.
\end{equation}
Similarly,
\begin{align}
E_2
&=\langle e^{M^\dagger\Delta t}a_j^\dagger a_j e^{M\Delta t}\rangle_t
=\langle (1+\Delta t M^\dagger)a_j^\dagger a_j(1+\Delta t M)\rangle_t+O(\Delta t^2) \nonumber\\
&=\langle a_j^\dagger a_j\rangle_t+\Delta t\langle M^\dagger a_j^\dagger a_j+a_j^\dagger a_j M\rangle_t+O(\Delta t^2)
= (V_j(t) + |z_j|^2) +\Delta t A_2+O(\Delta t^2),
\end{align}
with
\begin{equation}
A_2:=\langle M^\dagger a_j^\dagger a_j+a_j^\dagger a_j M\rangle_t.
\end{equation}
For the normalization factor,
\begin{align}
D
&=\langle e^{M^\dagger\Delta t}e^{M\Delta t}\rangle_t
=\langle (1+\Delta t M^\dagger)(1+\Delta t M)\rangle_t+O(\Delta t^2) \nonumber\\
&=1+\Delta t(\langle M\rangle_t+\langle M^\dagger\rangle_t)+O(\Delta t^2)
=1+\Delta t S+O(\Delta t^2),
\end{align}
where
\begin{equation}
S:=\langle M\rangle_t+\langle M^\dagger\rangle_t.
\end{equation}

Collecting the three expansions,
\[
E_1=z_j+\Delta t A_1+O(\Delta t^2),\qquad
E_2=|z_j|^2+V_j(t)+\Delta t A_2+O(\Delta t^2),\qquad
D=1+\Delta t S+O(\Delta t^2).
\]

The coefficients are obtained from coherent–state matrix elements of $M=\sum_k a_k^\dagger F_k(a)$:
\[
A_1=F_j(z)+z_j\Sigma(z)+z_j\Sigma(z)^*,\qquad
A_2=|z_j|^2(\Sigma+\Sigma^*)+z_j^*F_j(z)+z_j F_j(z)^*,\qquad
S=\Sigma+\Sigma^*,
\]
where
\[
\Sigma(z):=\sum_k z_k^*F_k(z).
\]

For notational simplicity, let $z\equiv z_j$ and $\Sigma\equiv\Sigma(z)$.  
Using these definitions,
\[
A_2+S|z|^2-(z^*A_1+z\,A_1^*)=0,
\]
as shown earlier. Hence the numerator obeys
\begin{equation}
\begin{aligned}
E_2D-|E_1|^2
&=\Big(\langle a_j^\dagger a_j\rangle_t-|\langle a_j\rangle_t|^2\Big)
+\Delta t\Big(A_2+S\langle a_j^\dagger a_j\rangle_t-(z^*A_1+zA_1^*)\Big)
+O(\Delta t^2)\\
&=\Var_j(t)+\Delta t\Big(A_2+S|z|^2-(z^*A_1+zA_1^*)\Big)
+\Delta t\,S\,\Var_j(t)+O(\Delta t^2)\\
&=\Var_j(t)+\Delta t\,S\,\Var_j(t)+O(\Delta t^2)\\
&=\Var_j(t)\big(1+\Delta t\,S\big)+O(\Delta t^2).
\end{aligned}
\end{equation}

Since $D^2=(1+\Delta t S)^2=1+2\Delta t S+O(\Delta t^2)$, we obtain the variance update
\begin{equation}
\begin{aligned}
\Var_j(t+\Delta t)
&= \frac{E_2D-|E_1|^2}{D^2} = \frac{\Var_j(t)\big(1+\Delta t S\big)+O(\Delta t^2)}{(1+\Delta t S)^2} \\[1em]
&= \frac{\Var_j(t)\big(1+\Delta t S\big)}{(1+\Delta t S)^2} + \frac{O(\Delta t^2)}{(1+\Delta t S)^2} \\[1em]
&= \frac{\Var_j(t)}{1+\Delta t S} + O(\Delta t^2) \\[1em]
&= \Var_j(t) \Big( 1 - \Delta t S + O(\Delta t^2) \Big) + O(\Delta t^2) \\[1em]
&= \bigl(1-2\Delta t\,\Re\Sigma(z(t))\bigr)\,\Var_j(t)+O(\Delta t^2).
\end{aligned}
\end{equation}

\noindent
In particular, if $\Var_j(t)=0$ (e.g., a product of coherent states at time $t$), then $\Var_j(t+\Delta t)=O(\Delta t^2)$.  
More generally, we can write:
\begin{equation}
\Var_j(t+\Delta t)=\bigl(1-2\Delta t\,\Re\Sigma(z(t))\bigr)\,\Var_j(t)+O(\Delta t^2),\ 
\text{and if }\Var_j(t)=0,\ \ \Var_j(t+\Delta t)=O(\Delta t^2).
\end{equation}

\subsection{1D Burgers' Simulation}
\label{app:burgers}

Consider the semi-discrete 1D viscous Burgers update with Reynolds number $R_e$
\begin{equation}
\frac{\partial u}{\partial t}
=
-u\frac{\partial u}{\partial x}
+
\frac{1}{R_e}\frac{\partial^2 u}{\partial x^2}.
\label{eq:burgers-pde}
\end{equation}
For the semi-discrete 1D viscous Burgers step with  grid spacing $\Delta x$, the stencil is
\begin{equation}
F_k(z)=\frac{z_{k+1}-2z_k+z_{k-1}}{R_e\,\Delta x^{2}}
-\frac{z_k}{2\Delta x}\bigl(z_{k+1}-z_{k-1}\bigr).
\label{eq:burgers-stencil}
\end{equation}
The first term diffusive contribution corresponds to the linear local operator which is nearest-neighbor in $d{=}1$.  
Thus $|E|\approx 2L$, $N=4L$, and the per-step depth is $\mathcal O(\log L)$.  
We use the generator of the form
\begin{equation}
A \;=\; \sum_{k} a_k^\dagger\,F_k(a),\qquad
F_k(a)
= \underbrace{\frac{a_{k+1}-2a_k+a_{k-1}}{R_e\,\Delta x^{2}}}_{\text{diffusion}}
\;-\;
\underbrace{\frac{a_k}{2\Delta x}\bigl(a_{k+1}-a_{k-1}\bigr)}_{\text{convection}}.
\label{eq:M-Burgers}
\end{equation}
For a product coherent state $\ket{z}$ at time $t$, write $z_k:=\langle a_k\rangle_t$.
By normal ordering, $F_k(z)$ is obtained by substitution $a_\ell\mapsto z_\ell$:
\begin{equation}
F_k(z)
=\frac{z_{k+1}-2z_k+z_{k-1}}{R_e\,\Delta x^{2}}
-\frac{z_k}{2\Delta x}\bigl(z_{k+1}-z_{k-1}\bigr).
\label{eq:Fk-burgers}
\end{equation}
The overlap $\Sigma(z):=\sum_\ell z_\ell^*F_\ell(z)$ then splits into diffusion and convection pieces,
\begin{equation}
\Sigma(z)=\Sigma_{\mathrm{diff}}(z)+\Sigma_{\mathrm{conv}}(z)
=\sum_k \frac{z_k^*(z_{k+1}-2z_k+z_{k-1})}{R_e\,\Delta x^2}
-\sum_k \frac{|z_k|^2}{2\Delta x}\,(z_{k+1}-z_{k-1}).
\label{eq:sigma-burgers}
\end{equation}

\paragraph{Forward-Euler mean update.}
By Eq.~\eqref{eq:euler-step} we have, for each grid point $k$,
\begin{equation}
z_k(t+\Delta t)=z_k(t)+F_k\big(z(t)\big)\,\Delta t+O(\Delta t^2),
\qquad
F_k(z)\ \text{given by \eqref{eq:Fk-burgers}}.
\label{eq:burgers-euler}
\end{equation}

\paragraph{First order variance.}
Specializing the first-order coefficients in Sec.~\ref{app:singlemode} to the Burgers generator \eqref{eq:M-Burgers} gives
\begin{equation}
A_1^{(k)} \,=\, \langle M^\dagger a_k + a_k M\rangle_t
= F_k(z)+z_k\,\Sigma(z)+z_k\,\Sigma(z)^*,
\qquad
S \,=\, \langle M\rangle_t+\langle M^\dagger\rangle_t \,=\, \Sigma(z)+\Sigma(z)^*,
\end{equation}
and
\begin{equation}
A_2^{(k)} \,=\, \langle M^\dagger a_k^\dagger a_k + a_k^\dagger a_k M\rangle_t
= |z_k|^2 \big(\Sigma(z)+\Sigma(z)^*\big)
+ z_k^* F_k(z)+ z_k F_k(z)^* .
\end{equation}
Inserting these into $V_k(t+\Delta t)=\frac{E_2D-|E_1|^2}{D^2}$ with the expansions
$E_1=z_k+\Delta t A_1^{(k)}+O(\Delta t^2)$, $E_2= |z_k|^2 + V_k(t)+\Delta t A_2^{(k)}+O(\Delta t^2)$,
$D=1+\Delta t S+O(\Delta t^2)$, the $O(\Delta t)$ term cancels identically (Sec.~\ref{app:singlemode}),
and therefore
\begin{equation}
\Var(a_k)_{\,t+\Delta t}= \Var(a_k)_{\,t}+O(\Delta t^2).
\label{eq:burgers-var}
\end{equation}
In particular, starting from a product of coherent states, $\Var(a_k)_{\,t}=0$ remains zero to first order.

\paragraph{Analysis of Numerical Stability.}
Write forward differences $\delta^+ z_k:=z_{k+1}-z_k$ and the discrete Laplacian $\Delta z_k:=z_{k+1}-2z_k+z_{k-1}=\delta^+ z_k-\delta^+ z_{k-1}$.
Under periodic boundary conditions, a discrete summation by parts gives
\begin{equation}
\sum_k z_k^*\,\Delta z_k
=\sum_k z_k^*(\delta^+ z_k-\delta^+ z_{k-1})
=\sum_k z_k^*\,\delta^+ z_k-\sum_k z_{k+1}^*\,\delta^+ z_k
=-\sum_k \delta^+ z_k^{\,*}\,\delta^+ z_k.
=-\sum_k \bigl|\delta^+ z_k\bigr|^2
\label{eq:disc-ip}
\end{equation}
Hence the diffusive part satisfies
\begin{equation}
\Re\,\Sigma_{\mathrm{diff}}(z)
=\frac{1}{R_e\,\Delta x^2}\,\Re\sum_k z_k^*\,\Delta z_k
=-\frac{1}{R_e\,\Delta x^2}\sum_k \bigl|\delta^+ z_k\bigr|^2
\;\le\; 0,
\label{eq:sigma-diff-neg}
\end{equation}
which is strictly dissipative (negative semidefinite).

For the convection contribution,
\begin{equation}
\Sigma_{\mathrm{conv}}(z)=-\frac{1}{2\Delta x}\sum_k |z_k|^2\,(z_{k+1}-z_{k-1}).
\end{equation}
If the coherent amplitudes are chosen globally real (fixing a global phase), $\Sigma_{\mathrm{conv}}(z)$ is real but does not have a definite sign in general. Nevertheless it can be bounded in terms of the discrete gradient:
\begin{equation}
\big|\Sigma_{\mathrm{conv}}(z)\big|
\;\le\;\frac{1}{2\Delta x}\sum_k |z_k|^2\big(|\delta^+ z_k|+|\delta^+ z_{k-1}|\big)
\;\le\;\frac{\|z\|_\infty^2}{\Delta x}\sum_k |\delta^+ z_k|
\;\le\;\frac{\epsilon}{R_e\,\Delta x^2}\sum_k |\delta^+ z_k|^2
+\frac{R_e\,\Delta x^2}{4\epsilon}\,\|z\|_\infty^4\,L,
\label{eq:conv-bound}
\end{equation}
for any $\epsilon>0$ (by Young’s inequality), where $L$ is the number of grid points and $\|z\|_\infty:=\max_k |z_k|$.
Combining \eqref{eq:sigma-diff-neg}–\eqref{eq:conv-bound} shows that for sufficiently diffusive regimes (small $R_e$ or adequately resolved grids), the dissipative term dominates $\Re\Sigma(z)$, ensuring stability of the normalized update; in practice this manifests as a controlled denominator $1+2\Delta t\,\Re\Sigma(z)$ in Eq.~\eqref{eq:ratio}.

\paragraph{Summary for the Burgers step.}
Given $F_k(z)$ in \eqref{eq:Fk-burgers} and $\Sigma(z)$ in \eqref{eq:sigma-burgers}, the one-step evolution obeys
\begin{equation}
z_k(t+\Delta t)=z_k(t)+\Delta t\,F_k(z(t))+O(\Delta t^2),\qquad
\Var(a_k)_{\,t+\Delta t}=\Var(a_k)_{\,t}+O(\Delta t^2),
\end{equation}
with the diffusive part contributing a strictly nonpositive $\Re\Sigma_{\mathrm{diff}}(z)$ by \eqref{eq:sigma-diff-neg}, and the convective part controlled by \eqref{eq:conv-bound}. 
In the common gauge where all $z_k$ are real, the identity $\sum_k z_k\,\Delta z_k=-\sum_k(\delta^+ z_k)^2\le 0$ makes the stabilizing role of diffusion explicit.

\paragraph{Numerical results for the Burgers' equation.}
For each grid point $j$, define the coherent amplitude as $z_j(t):=\Tr[a_j\rho(t)]$
and the variance
$\Var_j(t):=\Var(a_j)_t=\langle a_j^\dagger a_j\rangle_t-\big|\langle a_j\rangle_t\big|^2.$
With the short-time generator expressed in normal order as $A=\sum_k a_k^\dagger F_k(a)$,
the coherent-state algebra (Supplementary Materials~\ref{app:singlemode}) yields the following update formulae
\begin{align}
z_j(t+\Delta t)&=z_j(t)+\Delta t\,F_j(z)+\mathcal{O}(\Delta t^2), 
\label{eq:mean-burgers-main}\\
\Var_j(t+\Delta t)&=\Var_j(t)+\mathcal{O}(\Delta t^2).
\label{eq:var-burgers-main}
\end{align}
The cancellation of all $\mathcal{O}(\Delta t)$ contributions in Eq.~\eqref{eq:var-burgers-main} is nontrivial. Substituting the coherent-state expansions for the first and second moments into $\Var_j=(E_2D-|E_1|^2)/D^2$ shows that the terms proportional to $\Delta t$ cancel exactly (Supplementary Materials~\ref{eq:burgers-var}), leaving only $\mathcal{O}(\Delta t^2)$ corrections.  

For reference, we use the continuous 1D viscous Burgers' equation as shown in Eq.~\eqref{eq:burgers-pde} with its stencil, Eq.~\eqref{eq:burgers-stencil}. Under this discretization, the coherent amplitudes advance by the update in Eq.~\eqref{eq:mean-burgers-main}, reproducing nonlinear steepening and diffusion, while variances remain constant to first order.

Fig.~\ref{fig:burgers-mc-validation} illustrates the temporal evolution of the mean profile and its associated bias. The initial distribution travels to the right,
becoming asymmetric due to nonlinearity and developing a shock, while the error remains small. Table~\ref{tab:burgers-var} demonstrates the  estimates of the variance from $N=10^4$ samples per grid point match the analytic prediction within the expected $\sqrt{2/(N-1)}$ envelope.

\subsection{Extension to Two-Dimensional Lattices}
\label{app:2D}

The generator of the short-time dynamics now takes the form:
\begin{equation}
A = \sum_{p',q'} a_{p',q'}^\dagger F_{p',q'}(a),
\end{equation}
where $F_{p',q'}(a)$ is a function of all annihilation operators $\{a_{p',q'}\}$ on the lattice. The overlap scalar $\Sigma(z)$ is also a double summation:
\begin{equation}
\Sigma(z) := \sum_{p',q'} z_{p',q'}^* F_{p',q'}(z).
\end{equation}

\paragraph{Mean-Field Evolution.}
To find the first-order evolution of the mean field $z_{p,q}$, we expand the expectation value of the operator $O=a_{p,q}$:
\begin{equation}
\langle a_{p,q} \rangle_{t+\Delta t}
= \frac{\bra{z}(1+M^\dagger\Delta t)\,a_{p,q}\,(1+M\Delta t)\ket{z}}
{\bra{z}(1+M^\dagger\Delta t)(1+M\Delta t)\ket{z}}
+ O(\Delta t^2).
\end{equation}
The algebraic structure of the operator commutation relations is unchanged, so the coherent-state matrix elements are direct analogues of the 1D case:
\begin{align}
\bra{z}a_{p,q} M\ket{z} &= F_{p,q}(z)+z_{p,q}\Sigma(z), \qquad
\bra{z}M^\dagger a_{p,q}\ket{z} = z_{p,q}\Sigma(z)^*, \\
\bra{z}M\ket{z} &= \Sigma(z), \qquad
\bra{z}M^\dagger\ket{z} = \Sigma(z)^*.
\label{eq:2D-matrx}
\end{align}
Thus the denominator is
\begin{equation}
\mathrm{Den} = 1+2\Delta t\,\Re\Sigma(z)+O(\Delta t^2),
\end{equation}
and the numerator is
\begin{equation}
\mathrm{Num}=z_{p,q}+\Delta t F_{p,q}(z)+2\Delta tz_{p,q}\Re\Sigma(z)+O(\Delta t^2).
\end{equation}
When combined, the terms involving $\Re\Sigma(z)$ cancel identically, leaving the forward Euler update:
\begin{equation}\label{eq:fisher2d-mean}
z_{p,q}(t+\Delta t)=z_{p,q}(t)+\Delta t\,F_{p,q}(z(t))+O(\Delta t^2).
\end{equation}

\paragraph{First-Order Variance Evolution.}
Consider the variance of the annihilation operator at grid point $(p,q)$:
\begin{equation}
\Var_{p,q}(t) := \Var(a_{p,q}) = \langle a_{p,q}^\dagger a_{p,q} \rangle_t - \big|\langle a_{p,q} \rangle_t\big|^2.
\end{equation}
The variance update is again
\[
\Var_{p,q}(t+\Delta t) = \frac{E_2D-|E_1|^2}{D^2},
\]
with $E_1$, $E_2$, $D$ defined analogously to the 1D case.  

The key identity
\begin{equation}
A^{(p,q)}_2 + S|z_{p,q}|^2 - \big(z_{p,q}^* A^{(p,q)}_1+z_{p,q}(A^{(p,q)}_1)^*\big)=0
\end{equation}
holds in 2D just as in 1D, since it is purely algebraic and does not depend on the lattice dimension. Therefore the numerator reduces to
\[
E_2D-|E_1|^2 = \Var_{p,q}(t)\,(1+\Delta t S)+O(\Delta t^2),
\]
and dividing by $D^2=(1+\Delta t S)^2$ gives
\begin{equation}\label{eq:fisher2d-var}
\Var_{p,q}(t+\Delta t)=\bigl(1-2\Delta t\,\Re\Sigma(z)\bigr)\Var_{p,q}(t)+O(\Delta t^2).
\end{equation}
In particular, if $\Var_{p,q}(t)=0$ (a product of coherent states), then $\Var_{p,q}(t+\Delta t)=O(\Delta t^2)$.

\subsubsection{2D Fisher–KPP Simulation}
\label{app:2D-fisher}

Consider the short-time generator on a 2D lattice
\begin{equation}
\label{eq:Fisher-generator}
A \;=\; \sum_{i,j} a^{\dagger}_{i,j}\,F_{i,j}(a),
\qquad
\Sigma(z)\;:=\;\sum_{i,j} z_{i,j}^{*}\,F_{i,j}(z),
\end{equation}
with the grid point-wise first-order update from Sec.~\ref{app:2D}:
\begin{equation}
z_{i,j}(t+\Delta t)=z_{i,j}(t)+\Delta t\,F_{i,j} \big(z(t)\big)+O(\Delta t^2).
\end{equation}
We take the (discrete) Fisher-type advection–diffusion–reaction form
\begin{equation}
\label{eq:Fisher-2D-Fa}
\begin{aligned}
F_{i,j}(a)
=\;&\underbrace{\frac{y_j}{2\Delta x}
 \left(
\frac{a_{i+1,j}}{\sqrt{x_{i+1}^{2}+y_{j}^{2}}}
-\frac{a_{i-1,j}}{\sqrt{x_{i-1}^{2}+y_{j}^{2}}}
\right)
-\frac{x_i}{2\Delta y}
 \left(
\frac{a_{i,j+1}}{\sqrt{x_{i}^{2}+y_{j+1}^{2}}}
-\frac{a_{i,j-1}}{\sqrt{x_{i}^{2}+y_{j-1}^{2}}}
\right)}_{\text{advection }(u\cdot\nabla a)} \\[6pt]
&+\;\underbrace{\frac{1}{\mathrm{Pe}}
\left(
\frac{a_{i+1,j}-2a_{i,j}+a_{i-1,j}}{\Delta x^{2}}
+
\frac{a_{i,j+1}-2a_{i,j}+a_{i,j-1}}{\Delta y^{2}}
\right)}_{\text{diffusion }(\kappa\,\Delta a)} \\[4pt]
&+\;\underbrace{\mathrm{Da}\,\bigl(a_{i,j}-a_{i,j}^{2}\bigr)}_{\text{reaction }(\mathrm{Da}\,(a-a^{2}))}.
\end{aligned}
\end{equation}
For a product coherent state at time \(t\), write \(z_{i,j}:=\langle a_{i,j}\rangle_t\). Using the normal-order rule,
\begin{equation}
\label{eq:Fisher-2D-Fz}
\begin{aligned}
F_{i,j}(z)
=\;&\frac{y_j}{2\Delta x}
 \left(
\frac{z_{i+1,j}}{\sqrt{x_{i+1}^{2}+y_{j}^{2}}}
-\frac{z_{i-1,j}}{\sqrt{x_{i-1}^{2}+y_{j}^{2}}}
\right)
-\frac{x_i}{2\Delta y}
 \left(
\frac{z_{i,j+1}}{\sqrt{x_{i}^{2}+y_{j+1}^{2}}}
-\frac{z_{i,j-1}}{\sqrt{x_{i}^{2}+y_{j-1}^{2}}}
\right) \\[6pt]
&+\;\frac{1}{\mathrm{Pe}}
\left(
\frac{z_{i+1,j}-2z_{i,j}+z_{i-1,j}}{\Delta x^{2}}
+
\frac{z_{i,j+1}-2z_{i,j}+z_{i,j-1}}{\Delta y^{2}}
\right)
+\;\mathrm{Da}\,\bigl(z_{i,j}-z_{i,j}^{2}\bigr).
\end{aligned}
\end{equation}

\paragraph{Forward-Euler mean update.}
From Sec.~\ref{app:2D}, the coherent-state matrix elements lead to the same algebraic cancellation as in 1D, yielding
\begin{equation}
z_{i,j}(t+\Delta t)
= z_{i,j}(t) + \Delta t\,F_{i,j} \big(z(t)\big)+O(\Delta t^2),
\qquad
F_{i,j}(z)\ \text{given by }\eqref{eq:Fisher-2D-Fz}.
\end{equation}

\paragraph{First order variance}
Let \(\Var_{i,j}(t):=\Var(a_{i,j})=\langle a_{i,j}^\dagger a_{i,j}\rangle_t-|\langle a_{i,j}\rangle_t|^2\).
Exactly the same dimension-independent algebra as in Sec.~\ref{app:2D} shows that
\begin{equation}
\Var_{i,j}(t+\Delta t)
= \bigl(1-2\Delta t\,\Re\Sigma(z(t))\bigr)\,\Var_{i,j}(t)
+O(\Delta t^2),
\quad
\text{hence if }\Var_{i,j}(t)=0,\; \Var_{i,j}(t+\Delta t)=O(\Delta t^2),
\end{equation}
where \(\Sigma(z)=\sum_{i,j} z_{i,j}^* F_{i,j}(z)\).

\paragraph{Analysis of Numerical Stability.}
Write
\(\Sigma(z)=\Sigma_{\mathrm{adv}}(z)+\Sigma_{\mathrm{diff}}(z)+\Sigma_{\mathrm{reac}}(z)\).
Introduce forward differences
\(
\delta_x^+ z_{i,j}:=z_{i+1,j}-z_{i,j},\;
\delta_y^+ z_{i,j}:=z_{i,j+1}-z_{i,j}
\),
and the discrete Laplacian
\(
\Delta_h z_{i,j}:=\frac{\delta_x^+ z_{i,j}-\delta_x^+ z_{i-1,j}}{\Delta x^2}
+\frac{\delta_y^+ z_{i,j}-\delta_y^+ z_{i,j-1}}{\Delta y^2}.
\)
Under periodic (or homogeneous) boundary conditions, discrete summation by parts in 2D gives
\begin{equation}
\label{eq:2D-disc-ip}
\sum_{i,j} z_{i,j}^{*}\,\Delta_h z_{i,j}
= -\sum_{i,j} \left(\frac{|\delta_x^+ z_{i,j}|^2}{\Delta x^2}
+\frac{|\delta_y^+ z_{i,j}|^2}{\Delta y^2}\right),
\end{equation}
hence the diffusion contribution is strictly dissipative:
\begin{equation}
\label{eq:Sigma-diff-2D}
\Re\,\Sigma_{\mathrm{diff}}(z)
=\frac{1}{\mathrm{Pe}}\,\Re\sum_{i,j} z_{i,j}^{*}\,\Delta_h z_{i,j}
= -\frac{1}{\mathrm{Pe}}\sum_{i,j} \left(\frac{|\delta_x^+ z_{i,j}|^2}{\Delta x^2}
+\frac{|\delta_y^+ z_{i,j}|^2}{\Delta y^2}\right)
\;\le\;0.
\end{equation}
For the advection part (real if \(z\) is globally real but not sign-definite in general), using Young’s inequality with any \(\epsilon>0\), and denoting
\(
\|z\|_{\infty}:=\max_{i,j}|z_{i,j}|,
\;
\|w\|_{1}:=\sum_{i,j}|w_{i,j}|
\),
one obtains
\begin{equation}
\label{eq:Sigma-adv-bound}
\big|\Sigma_{\mathrm{adv}}(z)\big|
\;\le\;C_u
\left(\frac{\|\delta_x^+ z\|_{1}}{\Delta x}+\frac{\|\delta_y^+ z\|_{1}}{\Delta y}\right)
\;\le\;\frac{\epsilon}{\mathrm{Pe}} \sum_{i,j} \left(\frac{|\delta_x^+ z_{i,j}|^2}{\Delta x^2}
+\frac{|\delta_y^+ z_{i,j}|^2}{\Delta y^2}\right)
+\frac{C_u^2\,\mathrm{Pe}}{4\epsilon}\,N_{xy},
\end{equation}
where
\(
C_u:=\frac{1}{2}\max_{i,j} \left(\frac{|y_j|}{\sqrt{x_i^2+y_j^2}}+\frac{|x_i|}{\sqrt{x_i^2+y_j^2}}\right)
\)
bounds the discrete velocity prefactor and \(N_{xy}\) is the number of grid points.
For the reaction part,
\begin{equation}
\label{eq:Sigma-reac}
\Sigma_{\mathrm{reac}}(z)
= \mathrm{Da}\sum_{i,j} \bigl(z_{i,j}^{*}z_{i,j}-z_{i,j}^{*}z_{i,j}^{2}\bigr).
\end{equation}
If \(z_{i,j}\in\mathbb{R}\) (global phase fixed), then
\(
\Sigma_{\mathrm{reac}}(z)=\mathrm{Da}\sum_{i,j} \big(z_{i,j}^{2}-z_{i,j}^{3}\big),
\)
which is positive for small amplitudes (\(z\ll1\), linear growth) and becomes negative for large \(z\) (nonlinear saturation), reflecting the Fisher–KPP mechanism. For complex \(z\), use
\(
\big|\sum z^{*}z^{2}\big|
\le \sum |z|^{3} \le \frac{2}{3}\sum |z|^{2}+\frac{1}{3}\sum |z|^{4}
\)
to bound \(\Re\,\Sigma_{\mathrm{reac}}\).

\paragraph{Summary for the Fisher-KPP step.}
Given \(F_{i,j}(z)\) in \eqref{eq:Fisher-2D-Fz} and \(\Sigma(z)\) as above, one step of the 2D Fisher update obeys
\begin{equation}
z_{i,j}(t+\Delta t)=z_{i,j}(t)+\Delta t\,F_{i,j}(z(t))+O(\Delta t^2),\qquad
\Var_{i,j}(t+\Delta t)=\bigl(1-2\Delta t\,\Re\Sigma(z(t))\bigr)\Var_{i,j}(t)+O(\Delta t^2),
\end{equation}
with \(\Re\,\Sigma_{\mathrm{diff}}(z)\le 0\) strictly dissipative by \eqref{eq:Sigma-diff-2D}, advection controlled by \eqref{eq:Sigma-adv-bound}, and reaction providing growth at small amplitude and saturation at large amplitude. In diffusive/saturated regimes the normalized denominator \(1+2\Delta t\,\Re\Sigma(z)\) remains well-conditioned, ensuring stable first-order evolution.

\section{Trotterization time-evolution algorithm for solving Lid-Driven Cavity Problem}
\label{appsec:lid driven}
Each discrete stream function and vorticity component at grid location $(i,j)$ is embedded into the continuous-variable framework by associating two coherent-state amplitudes, $z^{(\omega)}_{i,j}$ and $z^{(\psi)}_{i,j}$, corresponding to $\omega_{i,j}$ and $\psi_{i,j}$ respectively. 
Independent bosonic modes are assigned so that
\begin{equation}
    |z\rangle = \bigotimes_{i,j} |z^{(\omega)}_{i,j}\rangle \otimes |z^{(\phi)}_{i,j}\rangle,
\end{equation}
and the dynamical variables are recovered via Eq.~\eqref{eq:expval}.
The discrete Navier-Stokes evolution is lifted into the bosonic operator generator by replacing each field with the corresponding annihilation operators.
For instance, Eq.~\eqref{eq:ns_incompress} is encoded schematically as:
\begin{equation*}
    F^\psi_{i,j}(a)
    = \frac{1}{\Delta x^2}\left(a^{\psi}_{i+1,j} - 2a^{\psi}_{i,j} + a^{\psi}_{i-1,j}\right) 
    + \frac{1}{\Delta y^2}\left(a^{\psi}_{i,j+1} - 2a^{\psi}_{i,j} + a^{\psi}_{i,j-1}\right) + a^{\omega}_{i,j},
\end{equation*}
and will be time-evolved with the Hamiltonian-like operator $M^{\psi} = \sum_{i,j}a^{\psi^\dagger}_{i,j}F^\psi_{i,j}(a)$

{
\paragraph{Mean Update: Trotterization Methods} 
Unlike the derivation in 2D Fisher-KPP equation, we use a Trotterized (Suzuki–Trotter) update to compute the mean-field dynamics for lid-driven cavity problem as an alternative approach.
The finite-time update follows 
Eq.~\eqref{eq:ratio} and can be written as% YG: To be added
: 
\begin{equation}
  z_k(t+\Delta t)
 = \frac{\langle z(t)|e^{M^\dagger\Delta t}\,a_k\,e^{M\Delta t}|z(t)\rangle}
        {\langle z(t)|e^{M^\dagger\Delta t} e^{M\Delta t}|z(t)\rangle}.
\end{equation}
To act on $a_k$ it is convenient to move $a_k$ through $e^{M\Delta t}$ using the Baker--Campbell--Hausdorff (BCH)/Campbell lemma identity:
\begin{equation}\label{app:campbell}
  e^{-X}Ye^X = \sum_{n=0}^{\infty}\frac{[Y,X]_n}{n!},
\end{equation}
where $[Y,X]_n = [[\dots[[Y,X],X],\dots,X],X]$ is the $n-$times iterated commutator. Choosing $X= M\Delta t$ and $Y=a_k$ yields
\begin{equation}\label{app:moveak}
    a_k e^{M\Delta t} = e^{M\Delta t}(a_k +[a_k,M]\Delta t
    + \tfrac{1}{2}[[a_k,M],M]\Delta t^2 + \mathcal{O}(\Delta t^3)).
\end{equation}
Substituting Eq.~\eqref{app:moveak} into the numerator and canceling the common factor $e^{M^\dagger\Delta t}e^{M\Delta t}$ between the numerator and denominator gives, to second order
\begin{equation}\label{app:ratio}
    z_k(t+\Delta t) =
    \frac{\langle z(t)|a_k +[a_k,M]\Delta t + \tfrac{1}{2}[[a_k,M],M]\Delta t^2|z(t)\rangle}
                         {\langle z(t)|z(t)\rangle} 
                         + \mathcal{O}(\Delta t^3).
\end{equation}
The first commutator reduces to
\begin{equation}\label{app:firstcomm}
  [a_k,M] = \sum_{j}[a_k,a_j^\dagger]\,F_j(a)= F_k(a),
\end{equation}
which shows that the linear term in $\Delta t$ produces $F_k(a)$.
For the quadratic term, the identity
\begin{equation}\label{app:commcalc}
    [[a_k,M],M] = [F_k(a),\sum_{j}a_j^\dagger F_j(a)]=\sum_{j}[F_k(a),a_j^\dagger]\,F_j(a),
\end{equation}
is needed. The commutator $[F_k(a),a_j^\dagger]$ is evaluated next. Let $F_k(a)$ be normally ordered and analytic, with power-series expansion $F_k(a)=\sum_{\boldsymbol{\alpha}} c_{\boldsymbol{\alpha}}\prod_\ell a_\ell^{\alpha_\ell}$. Then,
\begin{equation}\label{app:derivative}
  [F_k(a),a_j^\dagger]
  = \sum_{\boldsymbol{\alpha}} c_{\boldsymbol{\alpha}}\,\alpha_j\,a_j^{\alpha_j-1}\prod_{\ell\neq j}\! a_\ell^{\alpha_\ell}
  = \frac{\partial F_k}{\partial a_j}(a).
\end{equation}
This follows by linearity and $[a_\ell, a_j^\dagger]=\delta_{\ell j}$. Using Eq.~\eqref{app:derivative} in Eq.~\eqref{app:commcalc} gives
\begin{equation}\label{app:secondcomm}
  [[a_k,M],M]= \sum_{j}\frac{\partial F_k}{\partial a_j}(a)\,F_j(a).
\end{equation}
Inserting Eqs.~\eqref{app:firstcomm} and \eqref{app:secondcomm} into Eq.~\eqref{app:ratio} and applying the coherent-state expectation value yields
\begin{equation}\label{eq:euler-step-trotter}
    z_k(t+\Delta t)= z_k(t) + F_k(z)\Delta t + \frac{1}{2}\sum_{j}\frac{\partial F_k}{\partial z_j}(z)\,F_j(z)\,\Delta t^2 + \mathcal{O}(\Delta t^3),
\end{equation}
which is precisely the second-order Taylor expansion of the classical flow map generated by $\dot{\bm z}=\bm F(\bm z)$. 
Truncation in first order recovers Eq.~\eqref{eq:euler-step}, i.e.\ the forward Euler update.
}

Now we consider the lid-driven cavity problem.
The system is governed by the incompressible Navier-Stokes equations, written in stream-function-vorticity formulation:
\begin{subequations}
    \begin{eqnarray}
        \frac{\partial \omega}{\partial t} + \frac{\partial \psi}{\partial y}\frac{\partial \omega}{\partial x} - \frac{\partial \psi}{\partial x}\frac{\partial \omega}{\partial y} &= \frac{1}{\mathrm{Re}}\nabla^2 \omega, \label{eq:ns_momentum} \\
        \nabla^2 \psi &= -\omega. \label{eq:ns_incompress}
    \end{eqnarray}
\end{subequations}
Here $\psi$, $\omega$, and the velocity field $\mathbf{u} = (u,v)$ are connected via
\begin{equation}
    u = \frac{\partial \psi}{\partial y}, \quad v = -\frac{\partial \psi}{\partial x}, \quad \omega = \frac{\partial v}{\partial x} - \frac{\partial u}{\partial y},
\end{equation}
and $\mathrm{Re}$ is the Reynolds number defined based on the velocity of the lid and the length of the cavity. The physical domain is the unit square $\Omega = [0,1]\times[0,1]$.
The no-slip boundary conditions are imposed on the three stationary walls $\mathbf{u}(x,0) = \mathbf{u}(0,y) = \mathbf{u}(1,y) = \mathbf{0}$, while the top lid ($y=1$) is driven with a unit horizontal velocity $\mathbf{u}(x,1) = (1,0)$. 
The spatial domain is discretized using a uniform finite-difference grid with $N_x\times N_y$ interior points. Standard second-order central differences are used to approximate spatial derivatives.
In each time step, $\omega$ is updated from Eq.~\eqref{eq:ns_momentum} by advancing in time according to the Eq.~\eqref{eq:euler-step-trotter} update rule.
% (same as Eq.~\eqref{eq:fisher2d-mean} but derived differently).
Afterwards, $\psi$ is updated from Eq.~\eqref{eq:ns_incompress} by time-evolving the equation
\begin{equation}
    \frac{\partial\psi}{\partial t} = \nabla^2 \psi + \omega,
\end{equation}
to reach the steady state.\\ 

\begin{figure}[t]
    \centering
    \includegraphics[width=.98\linewidth]{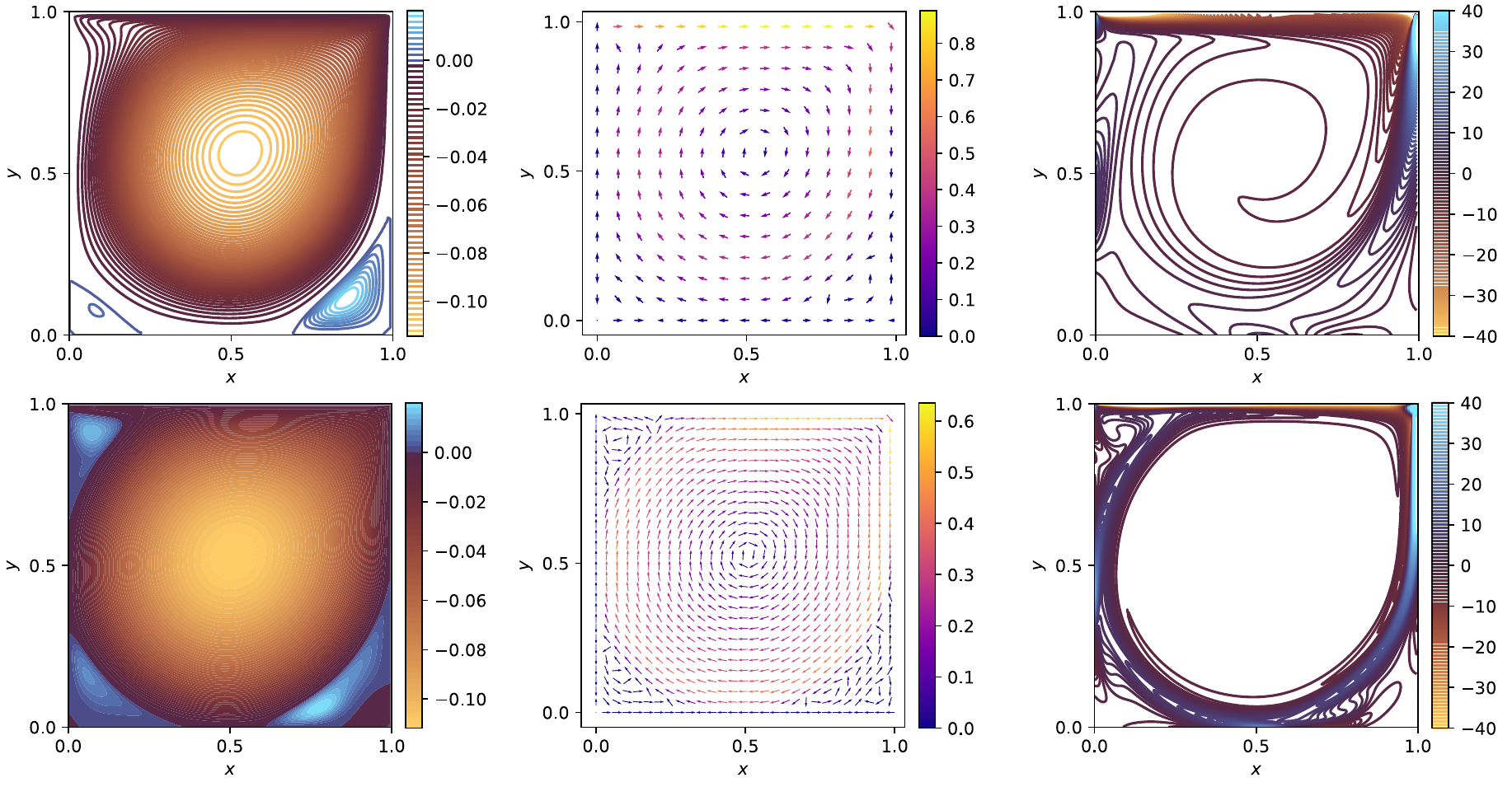}
    \caption{The lid-driven cavity simulation with a $128\times128$ grid and $\mathrm{Re} = 1000$ in top row and a $256\times256$ grid and $\mathrm{Re} = 10000$ in bottom row. All panels show the flow at the steady state. \textit{Left:} Contour plots of the stream function $\psi$. \textit{Middle:} The velocity field $\mathbf{u} = (u,v)$. \textit{Right:} The vorticity $\omega$. Simulation is done using bosonic simulator. The evolution follows the Euler update rule, and therefore, matches with DNS calculation.}
    \label{fig:fig2}
\end{figure}

Our numerical simulation shows that our methods can capture complex behavior under different Reynolds numbers.
The steady state solution of the lid-driven cavity is shown in Fig.~\ref{fig:fig2}. The initial condition is zero for all grid points. The simulations were performed in two setups: one with $N_x = N_y = 128$ grid and $\mathrm{Re} = 1,000$, and another with $N_x = N_y = 256$ grid and $\mathrm{Re} = 10,000$. The vorticity field $\omega$ advances in time with a time step of $\Delta t = 0.005$. The streamfunction $\psi$ advances with the time step $\Delta \tau = 5\times 10^{-6}$ until convergence. The simulations continue until $\|\omega(t+\Delta t) - \omega(t)\|_F \leq 10^{-5}$, where $\|.\|_F$ is the Frobenius norm. The solution field obtained is in agreement with the direct numerical simulation (DNS) in previous works 
% {\color{red}To Hirad: Is this " previous works" the DNS referred to in the legend for Fig. 4? Also, where is "DNS" defined?}
\cite{ghia1982high,kiffner2023tensor}, establishing the validity of the results.

\section{\label{sec:appe}Tensor Network Simulation of the Burgers' Equation}

\begin{figure}[t]
    \centering
    \includegraphics[width=0.5\linewidth]{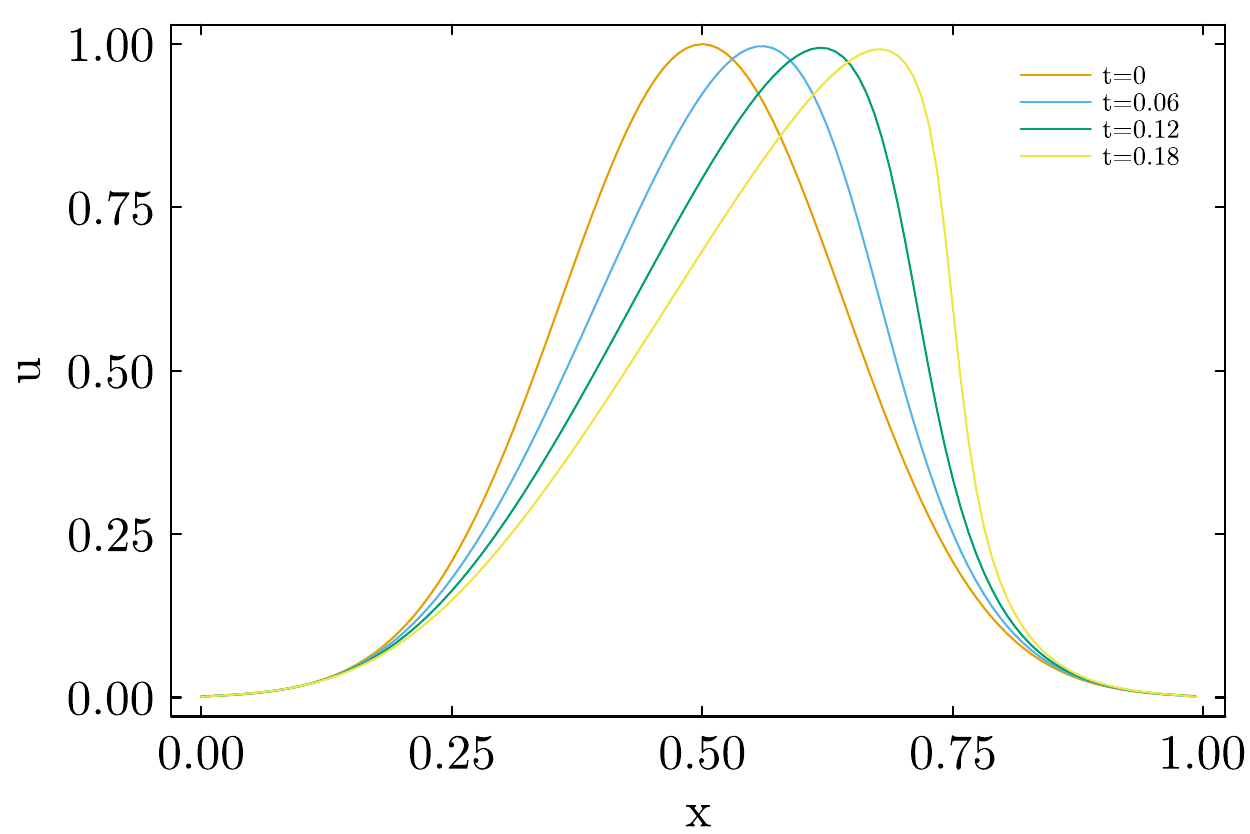}
    \caption{Tensor network simulation of the Burgers' equation using a Trotterized TEBD time-evolution scheme. The initial condition is a Gaussian velocity profile, which evolves into a shock-like structure that is subsequently smoothed by viscosity. The resulting profiles at $t = 0, 0.06, 0.12,$ and $0.18$ are in excellent visual agreement with the reference solution shown in Fig.~\ref{fig:burgers-mc-validation}.}
    \label{fig:tnt-burgers}
\end{figure}

In order to assess the feasibility of the Trotterized time-evolution scheme used in this work, we simulate the viscous Burgers' equation on a one-dimensional lattice with $L = 128$ sites and Dirichlet boundary conditions at the two ends of the domain. The velocity field $u(x,t)$ is discretized as $u_j(t) \equiv u(x_j,t)$ at lattice points $x_j$, and its dynamics is approximated by Hamiltonian-like operator representing the finite-difference scheme. We employ the TEBD algorithm with a Trotterized time-evolution operator and a fixed time step of $\Delta t = 10^{-5}$. Each lattice site is represented by a local bosonic Hilbert space, truncated to a maximum photon occupation number of $n_{\max} = 5$. This cutoff corresponds to a local dimension of $d = n_{\max} + 1$ and was found to be the smallest value that yields numerically stable and visually accurate dynamics for the parameter regime considered. Figure~\ref{fig:tnt-burgers} shows the simulated profiles at times $t = 0, 0.06, 0.12,$ and $0.18$. Starting from a smooth Gaussian initial condition, the tensor network time evolution correctly captures both the steepening of the profile and the subsequent viscous smoothing, reproducing the shock-like structure and its broadening in time. Within plotting resolution, the TN results are indistinguishable by eye from the reference solution shown in Fig.~\ref{fig:burgers-mc-validation}, indicating that the Trotterized scheme is capable of faithfully reproducing Burgers' dynamics with a modest local truncation and time step.

\section{\label{sec:appb}Open-System Dynamics in the $P$-Representation: Detailed Derivation}

This section expands the derivation of the PDE that governs the Glauber–Sudarshan $P$-function and the resulting evolution of the coherent-state amplitudes in the presence of photon loss. The density operator is expanded as
\begin{equation}\label{appB:Pexpansion}
  \rho(t) = \int d^2 z\, P(z,z^*,t)\,|z\rangle\langle z|,\qquad
  d^2 z \equiv \prod_{j} d^2 z_j .
\end{equation}

The evolution of density matrix under the generator $M=\sum_j a_j^\dagger F_j(a)$ and photon loss at the rate $\gamma$ are governed by Eq.~\eqref{eq:mastereq}, which takes the form
\begin{equation}\label{appB:master}
    \frac{\partial \rho}{\partial t} = \sum_j\bigg[a_j^\dagger F_j(a)\rho + \rho F_j^\dagger(a) a_j
    + \gamma a_j\rho a_j^\dagger - \frac{\gamma}{2}a_j^\dagger a_j\rho - \frac{\gamma}{2}\rho a_j^\dagger a_j\bigg].
\end{equation}

For the effect of $a_j^\dagger$ and $a_j$ on the density matrix basis $|z\rangle\langle z|$, integration by parts for $a_j^\dagger |z\rangle\langle z|$ and $|z\rangle\langle z| a_j$ yields the convenient differential actions
\begin{align}
a_j^\dagger \ket{z}\bra{z}
  &= \Bigl(-\partial_{z_j}+ z_j^*\Bigr)\ket{z}\bra{z}, \label{appB:aid} \\
\ket{z}\bra{z}\, a_j
  &= \Bigl(-\partial_{z_j^*}+ z_j\Bigr)\ket{z}\bra{z}. \label{appB:aconj}
\end{align}

when the derivatives are understood to act on everything to their right inside the $P$-integral. From moving $\partial/\partial z_j$ and $\partial/\partial z_j^*$ off the kets/bras onto $P$ by parts, boundary terms vanish under the standard assumption that $P$ decays sufficiently rapidly at $|z|\to\infty$. Using Eqs.~\eqref{appB:aid}–\eqref{appB:aconj} and normal ordering of $F_j(a)$, the coherent and dissipative components in Eq.~\eqref{appB:master} are obtained as
\begin{equation} \label{appB:Mrho}
    M\rho
  = \int d^2 z\, P(z,z^*,t)\sum_j F_j(z)\, a_j^\dagger |z\rangle\langle z| 
  = \int d^2 z \sum_j\!\left(-\frac{\partial}{\partial z_j} + z_j^*\right)\!\big[ P(z,z^*,t)\,F_j(z)\big]\,|z\rangle\langle z|,
\end{equation}
\begin{equation} \label{appB:rhoMd}
    \rho M^\dagger
  = \int d^2 z\, P(z,z^*,t)\sum_j F_j(z^*)\, |z\rangle\langle z| a_j 
  = \int d^2 z \sum_j\!\left(-\frac{\partial}{\partial z_j^*} + z_j\right)\!\big[ P(z,z^*,t)\,F_j(z^*)\big]\,|z\rangle\langle z|,
\end{equation}

\begin{equation} \label{appB:gain}
  \sum_j \gamma a_j \rho a_j^\dagger
  = \gamma \int d^2 z\, P(z,z^*,t) \sum_j |z_j|^2\,|z\rangle\langle z|,
\end{equation}
\begin{equation} \label{appB:loss}
  -\frac{\gamma}{2}\sum_j \big(a_j^\dagger a_j\rho + \rho a_j^\dagger a_j\big)
  = 
  -\frac{\gamma}{2}\int d^2 z \sum_j \Bigg[\Big(-\frac{\partial}{\partial z_j} + z_j^*\Big)\!\big(P z_j\big) 
  + \Big(-\frac{\partial}{\partial z_j^*} + z_j\Big)\!\big(P z_j^*\big)\Bigg]\,|z\rangle\langle z|.
\end{equation}

Here and below, $P\equiv P(z,z^*,t)$ is used for compactness. Collecting Eqs.~\eqref{appB:Mrho}–\eqref{appB:loss} in Eq.~\eqref{appB:master} and comparing the coefficients of $|z\rangle\langle z|$ gives the first-order PDE
\begin{equation}\label{appB:PDE}
  \frac{\partial P}{\partial t}
  = \sum_j \Bigg[
    \frac{\partial}{\partial z_j}\bigg(\frac{\gamma}{2} z_j P - F_j(z)P\bigg)
    + \frac{\partial}{\partial z_j^*}\bigg(\frac{\gamma}{2} z_j^* P - F_j(z^*)P\bigg)
    + \big(z_j^* F_j(z) + z_j F_j(z^*)\big) P
  \Bigg].
\end{equation}

The last term arises from the $z_j^*$ and $z_j$ components in Eqs.~\eqref{appB:aid}–\eqref{appB:aconj}. This term reflects the change in the norms of coherent states. In practice, calculations are performed at each time step. Therefore, the norm can be restored to the coherent states so that the normalization of $\rho$ is preserved by Eq.~\eqref{appB:PDE}. The PDE can be viewed as a generalized drift equation for $P$, with photon loss adding a linear contraction in phase space. Let $\langle\!\langle g \rangle\!\rangle \equiv \int d^2 z\, g(z,z^*)\, P(z,z^*,t)$ denote $P$-averages. For the first moments $z_k$
\begin{equation}\label{appB:momentdef}
  \frac{d}{dt}\langle\!\langle z_k \rangle\!\rangle
  = \int d^2 z\, z_k\, \frac{\partial P}{\partial t}.
\end{equation}

Note that the first moment can also be seen as the expectation values of the annihilation operator, namely
\begin{equation}
    \langle a_k\rangle = \text{Tr}(a_k\rho) = \int d^2 z\, P(z,z^*,t)z_k\text{Tr}(|z\rangle\langle z|) 
    = \int d^2 z\, P(z,z^*,t)z_k
    = \langle\!\langle z_k \rangle\!\rangle.
\end{equation}

Insertion of Eq.~\eqref{appB:PDE} and integration by parts (with vanishing boundary terms) give
\begin{equation} \label{appB:momentraw}
  \frac{d}{dt}\langle\!\langle z_k \rangle\!\rangle
  = -\left\langle\!\!\left\langle \frac{\gamma}{2} z_k - F_k(z)\right\rangle\!\!\right\rangle
  + \sum_j \left\langle\!\!\left\langle z_r z_j^* F_j(z) + z_r z_j F_j(z^*)\right\rangle\!\!\right\rangle .
\end{equation}

For sharply peaked $P$ (Gaussian-like packet centered at $z$ with small covariance), the factorization
\begin{equation}\label{appB:mfclosure}
  \langle\!\langle G(a,a^\dagger)\rangle\!\rangle \approx G\big(\langle\!\langle z\rangle\!\rangle,\langle\!\langle z^*\rangle\!\rangle\big),
\end{equation}
holds for normally ordered $G$, where $G$ denotes any normally ordered polynomial or analytic function of the mode operators (e.g. finite sums of monomials). In the Glauber–Sudarshan $P$-representation one has the exact mapping $\langle\!\langle G(a,a^\dagger)\rangle\!\rangle = \int d^2 z\, P(z,z^*,t)G(z,z^*)$ \cite{glauber1963coherent,sudarshan1963equivalence,walls2025quantum,GardinerZoller2004QuantumNoise,Carmichael1999SMQO1}. If $P$ is Gaussian-like and narrow with covariance $\Sigma$, a Taylor expansion about the mean $z$ shows the leading correction is $O(\|\Sigma\|)$, so evaluating $G$ at the means gives the standard mean-field (Gaussian) closure \cite{sudarshan1963equivalence,walls2025quantum,GardinerZoller2004QuantumNoise,Carmichael1999SMQO1,vanKampen2007StochasticProcesses,Haken1984LaserTheory,ScullyZubairy1997QuantumOptics}. Under this closure, Eq.~\eqref{appB:momentraw} reduces to
\begin{equation}\label{appB:meantraj}
  \frac{d z_k}{dt} = F_k(z) - \frac{\gamma}{2}\, z_k,
\end{equation}
where $z_k \equiv \langle\!\langle z_k \rangle\!\rangle = \langle a_k \rangle$. Equation~\eqref{appB:meantraj} reproduces the drift stated in the main text, implying an exponential decay of the coherent amplitude by $\exp(-\gamma t/2)$.

\end{document}